\documentclass[12pt,preprint]{aastex}
\usepackage{color}

\shorttitle{A Galaxy Merger Scenario for NGC~1550}
\shortauthors{Kawaharada et al.}

\begin{document}

\title{A Galaxy Merger Scenario for the NGC~1550 galaxy from Metal Distributions in the X-ray Emitting Plasma}

\author{
Madoka {\sc Kawaharada}\altaffilmark{1}, Kazuo {\sc Makishima}\altaffilmark{1,2}, Takao {\sc Kitaguchi}\altaffilmark{2}, Sho {\sc Okuyama}\altaffilmark{2}, \\ Kazuhiro {\sc Nakazawa}\altaffilmark{2}, Kyoko {\sc Matsushita}\altaffilmark{3},
{\sc and} Yasushi {\sc Fukazawa}\altaffilmark{4} 
}  

\altaffiltext{1}{Cosmic Radiation Laboratory, RIKEN, 2-1 Hirosawa, Wako, Saitama 351--0198; kawahard@crab.riken.jp.}
\altaffiltext{2}{Department of Physics, The University of Tokyo,
                7-3-1 Hongo,  Bunkyo-ku,  Tokyo 113--0033.}
\altaffiltext{3}{Department of Physics, Tokyo University of Science,
                 1-3 Kagurazaka, Shinjuku-ku, Tokyo 162-8601.}
\altaffiltext{4}{Department of Physical Science, Hiroshima University,
                1-3-1 Kagamiyama, Higashi-hiroshima 739-8526.}

\begin{abstract}
The elliptical galaxy NGC~1550 at a redshift of $z=0.01239$,
identified with an extended X-ray source RX~J0419+0225, was
observed with {\it XMM-Newton} for 31 ks. From the X-ray data
and archival near infra-red data of Two Micron All Sky survay, 
we derive the profiles of components constituting the NGC~1550 
system; the gas mass, total mass, metal mass, and galaxy luminosity. 
The metals (oxygen, silicon, and iron) are extended to $\sim 200$ kpc 
from the center, wherein $\sim$ 70\% of the $K$-band luminosity is 
carried by NGC~1550 itself.
As first revealed with {\it ASCA}, the data reconfirms the
presence of a dark halo, of which the mass 
($1.6 \times 10^{13}\; M_{\odot}$) is typical of a galaxy group
rather than of a single galaxy. Within 210 kpc, the $K$-band 
mass-to-light ratio reaches $75\; M_{\odot}/L_{\odot}$, 
which is comparable to those of clusters of galaxies.
The iron-mass-to-light ratio profile (silicon- and oxygen-mass-to-light 
ratio profiles as well) exhibits about two orders of magnitude decrease 
toward the center. Further studies comparing mass 
densities of metals with those of the other cluster components reveal 
that the iron (as well as silicon) in the ICM traces very well the total 
gravitating mass, whereas the stellar component is significantly more 
concentrated to within several tens kpc of the NGC~1550 nucleus. 
Thus, in the central region, the amount of metals is significantly 
depleted for the luminous galaxy light. Among a few possible 
explanations of this effect, the most likely scenario is that
galaxies in this system were initially much more extended than today,
and gradually fell to the center and merged into NGC~1550.

\end{abstract}


%%% 1. Introduction %%%%%%%%%
\section{Introduction}
%%%%%%%%%%%%%%%%%%%%%%%%%%%%%
Early-type galaxies, despite their optical homogeneity 
(Kodama \& Matsushita 2000), exhibit two distinct subclasses 
with respect to their X-ray properties 
(Matsushita 2001; Vikhlinin et al. 1999);
{\em X-ray extended} galaxies, and {\em X-ray compact} ones.
Compared to objects in the latter class, those in the former 
class are more luminous and extended in X-rays,
and are surrounded by group-scale potential halos 
(Matsushita et al. 1998; Kawaharada et al. 2003). 
Some of such X-ray extended ones are cD or XD (X-ray dominant)
galaxies in groups or clusters, while others appear optically isolated,
which we call {\em isolated X-ray extended galaxies} (IXEGs).

IXEGs are also called fossil groups (Jones et al. 2003),
or X-ray overluminous elliptical galaxies (Vikhlinin et al. 1999). 
The original definition of a fossil group (Jones et al. 2003) is that
its X-ray bolometric luminosity is higher than $4.8 \times 10^{41}$ 
erg s$^{-1}$ (for a Hubble constant of 72 km s$^{-1}$ Mpc$^{-1}$), 
and that the two brightest 
member galaxies within half the virial radius have an $R$-band 
magnitude difference ($\Delta m_{12}$) larger than 2. 
Jones et al. (2003) thus identified five fossil groups, and found 
that they have significantly higher X-ray luminosities than normal 
groups of similar total optical luminosities. These objects were 
thought to represent 8--20 \% of all systems with the X-ray 
luminosity of $\ge 4.8 \times 10^{41}$ erg s$^{-1}$.
Later, more than 30 objects have been identified as fossil group 
candidates (Santos et al. 2007) by combining the Sloan Digital 
Sky Survey (SDSS) data with those from the Rosat All Sky Survey 
(RASS). From a study of seven fossil groups, Khosroshahi et al. (2007) 
found that they show systematically higher X-ray luminosities 
and temperatures than normal galaxy groups of similar optical
luminosities, and higher mass concentrations. These authors
attributed the more enhanced X-ray emission from these fossil
groups to their cuspy potential shape. 

As suggested by cosmological simulations (D'Onghia et al. 2005; 
Dariush et al. 2007), a likely formation scenario of an IXEG
(or a fossil group) is that it was an ordinary galaxy group 
at the beginning, and its member galaxies have merged into a 
central galaxy which is now observed as an IXEG (e.g. Jones et al. 2003). 
Indeed, a rich population of dwarf galaxies, found around the 
IXEG NGC~1132 (Mulchaey \& Zabludoff 1999), provides good
evidende, because such dwarf members are expected to have
very long merging time scales, whereas the brightest group
members are thought to have merged in a few tenths of the
Hubble time if the galaxy density was initially high enough
(e.g. Barnes 1989; Zabludoff \& Mulchaey 1998). Nevertheless,
we may not be able to rule out an alternative possibility that
IXEGs have been one-galaxy systems since their formation, 
and their group-like properties are inherent to their birth process.  
In the present paper, we address this issue from X-ray observations
of one such object, NGC~1550.

Among IXEGs ever found, the S0 galaxy NGC~1550 is one of the 
nearest (at a redshift of $z=0.01239$) and the X-ray brightest. 
Historically, an extended ($\sim 14'$ in radius) X-ray emission 
named RX~J0419+0225 was first discovered by the RASS
from a position centered on this galaxy. Although the NGC~1550 system 
does not completely meet the condition of $\Delta m_{12} \geq 2$ for
fossil groups given by Jones et al. (2003), it has only a few other 
faint galaxies within the bright extended X-ray emitting region, 
and hence an ideal target to explore the nature of IXEGs.
In 1998, we observed this object with {\it ASCA} for 48 ks; 
assuming the Hubble constant as 
$H_0 = 72\; \rm{km\; s^{-1}\; Mpc^{-1}}$,
we found that the temperature (1.4 keV), X-ray luminosity
($5.1 \times 10^{42}\; \rm{erg\;s^{-1}}$),
gas mass ($3.4 \times 10^{11}\; M_{\odot}$), 
and the total mass ($1.6\times 10^{13}\; M_{\odot}$) 
of this object are typical for those of galaxy groups, 
rather than of single galaxies (Kawaharada et al. 2003; 
hereafter Paper I). In addition, its mass-to-light ($B$ band) ratio 
turned out to be as high as
$\simeq 380\; M_{\odot}/L_{\odot}$, comparable to those of
galaxy clusers. Thus, NGC~1550, associated with the X-ray source 
RX~J0419+0225, can be regarded as a typical IXEG, surrounded by 
a group-scale gravitational potential. 

With its high surface brightness in X-rays and its large X-ray
angular extent, NGC~1550 is expected to provide a new X-ray test
of the galaxy merger scenario. It is widely agreed that metals 
in the X-ray emitting plasmas (Intra-Cluster Medium, or ICM) in galaxy
groups and clusters have been supplied by member galaxies, either at an
early stage of the system formation, and/or over the Hubble time 
(e.g Matteucci \& Vettolani 1988).
Furthermore, the metals ejected to the ICM are not 
expected to diffuse significantly from the place where they were 
first deposited.
For example, 
an iron ion ejected into the ICM from a galaxy is slowed down
by collisions with surrounding protons and helium ions (Rephaeli 1978).
Since the diffusion constant of iron in the ICM is $\sim 10^{26}$ cm$^{-2}$ s 
(Ezawa et al. 1997; 2001), iron ions can diffuse at most $\sim 10$ kpc over
the Hubble time, assuming no bulk flows.
Then, if NGC~1550 is a one-galaxy system from the birth, its metals should
concentrate within the region of the central galaxy, $\sim 30$ kpc in radius.
Otherwise, the metals should extend over the group-scale potential well. 
Since the {\it ASCA} data were too limited in statistics to reliably
answer this inquiry, we resorted to {\it XMM-Newton} observations.
Our proposal based on this idea was accepted in the {\it XMM-Newton} 
AO-2 cycle, and the observation was carried out successfully in 2003.
The prime goal of the present study is to compare spatial distributions 
of metals in the ICM and the stellar mass. We utilize the
archival near infra-red data of Two Micron All Sky Survey (2MASS)
for the information of galaxy light. 

We report the {\it XMM-Newton} observation 
of NGC~1550 in \S2, followed by
the analysis and results of the EPIC data in \S3.
In \S4, we briefly evaluate the near infra-red data of NGC~1550 and
surrounding galaxies. The X-ray and near infra-red data are compared
in \S5, focusing in particular whether the metals in the ICM are
spatially more extended than the stellar light. In \S6, we discuss 
the nature of the NGC~1550 system based on these results.
Quantities quoted from the literature are rescaled to 
$H_0 = 72\; \rm{km\; s^{-1}\; Mpc^{-1}}$.
The definition of solar abundance is from Anders and Grevesse~(1988). 
Errors are given at the 90 \% confidence level.

%%% 2. Observation %%%%%%%%%%%%%%%%%%%%%%%%%%%%%%%%%%%%%%%
\section{Observation}
%%%%%%%%%%%%%%%%%%%%%%%%%%%%%%%%%%%%%%%%%%%%%%%%%%%%%%%%%%%%%%%%%%%%%

%--------- 2.1 ------------------------
\subsection{\it Data preparation}
%--------- 2.1 ------------------------
NGC~1550 was observed with {\it XMM-Newton} on 2003 February 22 for 31 ks, 
with the Prime Full Window mode and medium filter for the EPIC, 
and the High Event Rate with SES mode for the RGS. 
To create event files, we processed the Observation Data File of the 
PN and MOS data using ``epchain'' and ``emchain'' tasks of the standard 
analysis package (SAS) version 6.5.0, respectively.
The created event files were further filtered with respect to 
the event quality flag and pixel patterns. We chose events which have 
good quality (not collected at the edge of CCDs or bad pixels) 
and single pixel pattern for PN, 
and good quality and less than quadruple
pattern for MOS taking into account the $\sim 4$
times smaller MOS pixel size ($1.''1$ squared). 
In order to reject background flares, we made $10$--$12$ keV count-rate
histograms of each detector with 100 sec integration, 
and discarded those time bins which are deviated by more than 
2$\sigma$ from the mean rate.
After this event selection, the effective exposure became 19.6 ks (PN),
21.1 ks (MOS1), and 21.4 ks (MOS2). 
Point sources were excluded  by ``ewavlet'' task of the SAS 
with the detection threshold set at 7$\sigma$. Then, we made response 
files for each CCD through ``rmfgen'' and ``arfgen'' tasks.

%--------- 2.2 ------------------------
\subsection{\it Background estimate}
\begin{figure}
  \begin{center}
   \includegraphics[width=0.4\textwidth,angle=0,clip]{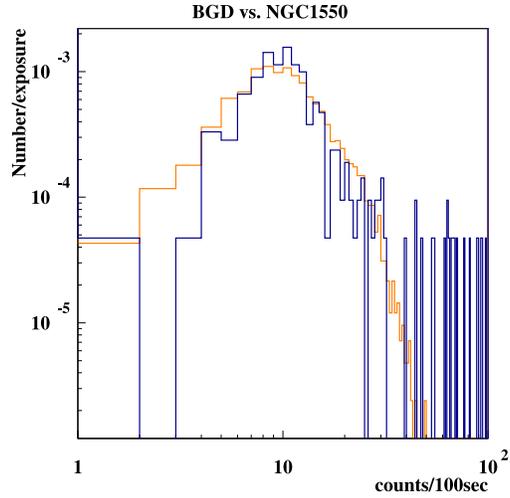} 
  \end{center}
  \caption{MOS1 $10-12$ keV count-rate histograms of the NGC~1550 observation (dark blue) and the template background (orange) with 100 sec integration, both normalized to their exposures.}
\label{fig:n1550_bgd_cor}
\end{figure}

%--------- 2.2 ------------------------
For the present purpose, an accurate background subtraction is 
of essential importance: 
in order to measure the ICM metallicity out to large radii where
the X-ray surface brightness becomes low, we need to accurately 
determine not only atomic lines, but also contnuum.
As template background files of the three 
EPIC detectors, we used those data taken with the medium filter, 
publicly released by the University of Birmingham group\footnote{http://www.sr.bham.ac.uk/xmm3} (Read and Ponman 2003).
These background files are superpositions of multiple observations, 
and hence have higher statistics and smaller systematical errors than
other background datasets currently available.
These background events were rearranged to match the observational
condition of NGC~1550, as originally developed with the {\it ASCA} GIS
(Ishisaki 1995) and also applied to the {\it XMM-Newton} EPIC data 
by Kaastra et al. (2004).
That is, we sorted the entire background data into subsets 
according to 100-sec averaged $10-12$ keV count rates, 
and created background spectra separately for individual subsets. 
We then combined them into a single
background spectrum (one for each detector), using appropriate 
weights which are specified by the $10-12$ keV count rate histograms
(averaged over 100 sec) of the actual NGC~1550 observation.
For example, Figure~\ref{fig:n1550_bgd_cor} is the
MOS1 $10-12$ keV count (per 100 sec) distribution during the
NGC~1550 observation, compared with that of the template background. 
Thus, the two distributions are reasonably matched with each other.
The background spectrum to be utilized in the present analysis, 
$B(E)$, is then synthesized as
\begin{eqnarray}
B(E) = \frac{\Sigma\; T_j B_j(E)}{\Sigma\; T_j}, 
\end{eqnarray}
where $E$ is the energy, $T_j$ is the total residence time when the
$10-12$ keV count is in the $j$-th interval, and $B_j(E)$ is 
the background template for the same $j$-th interval.

\begin{figure}
  \begin{center}
   \includegraphics[width=0.38\textwidth,angle=-90,clip]{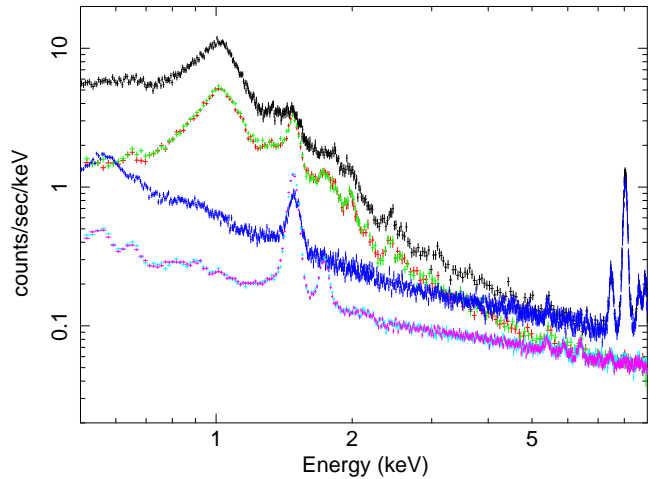} 
  \end{center}
  \caption{$0.5-9.0$ keV Raw spectra of PN (black), MOS1 (red), 
and MOS2 (green) extracted from the $0'-14'$ region. For comparison, 
estimated background spectra of PN (blue), MOS1 (cyan), and MOS2 (magenta) 
are also plotted. 
}
\label{fig:n1550_src_bgd}
\end{figure}

Katayama et al. (2004) studied the properties 
of the PN background and found that $2-7$ keV background rate 
correlates tightly with 
that in the $10-12$ keV band. Its correlation coefficient is
0.929, and the scatter around the correlation is less than 8\%.
Kaastra et al. (2004) showed that the MOS $0.2-10$ keV background 
flux scales proportionally to the $10-12$ keV background rate, 
and the spectral shape does not depend on the rate within 
an uncertaintly of 10\%. Therefore, the above method of 
background synthesis is considered to reproduce the background 
with an accuracy better than 10\%. Figure~\ref{fig:n1550_src_bgd} 
compares the $0.5-9.0$ keV raw EPIC spectra of NGC~1550 with the 
estimated background spectra. In the higher energy range above 
$\sim 6$ keV, the two spectra agree within $10$\% (1$\sigma$).

Since the Galactic absorption toward NGC~1550 is rather high as
$1.15 \times 10^{21}$ cm$^{-2}$ (Dickey \&  Lockman 1990),
the cosmic X-ray background (CXB) which dominates the low energy
band might be overestimated when the summed blank sky
background is used. However, at 0.6 keV, the count rate ratios
among the signal, particle background, and CXB
is estimated to be $1:0.2:0.06$. If the Galactic absorption
is $3 \times 10^{20}$ cm$^{-2}$, this ratio becomes $1:0.2:0.09$.
Therefore, the difference of the CXB level caused by using the
averaged blank sky background is at most $\sim 3$\% and $\sim 15$\%
of the signal and particle background at 0.6 keV, respectively.
In order to deal with this uncertainty
of cosmic X-ray background, and of the particle background as well,
we put systematic errors to the spectra, following Kaastra et al. (2004). 
The systematic errors in the source spectrum were set at 5\% in $0.5-2.0$ keV, 
and 10\% above $2.0$ keV. To the background spectrum, we assigned 
errors of 25\% in $0.5-0.7$ keV, 15\% between $0.7-2.0$ keV, 
and 10\% above 10 keV. 
We added these systematic errors in quadrature to the corresponding
statistical errors.

%%% 3. Results%%%%%%%%%%%%%%%%%%%%%%%%%%%%%%%%%%%%%%%
\section{X-ray Data Analysis and Results}
%%%%%%%%%%%%%%%%%%%%%%%%%%%%%%%%%%%%%%%%%%%%%%%%%%%%%

%--------- 3.1 ------------------------
\subsection{\it Radial surface brightness}
\begin{figure}
%  \begin{center}
    %\FigureFile(80mm,80mm){./fig/n1550_dss_xmm.ps}
    %\FigureFile(80mm,80mm){./fig/mkimage2.ps}
 \begin{minipage}{0.5\hsize}
  \begin{center}
\vspace{0.7cm}
\includegraphics[width=1.04\textwidth,angle=0,clip]{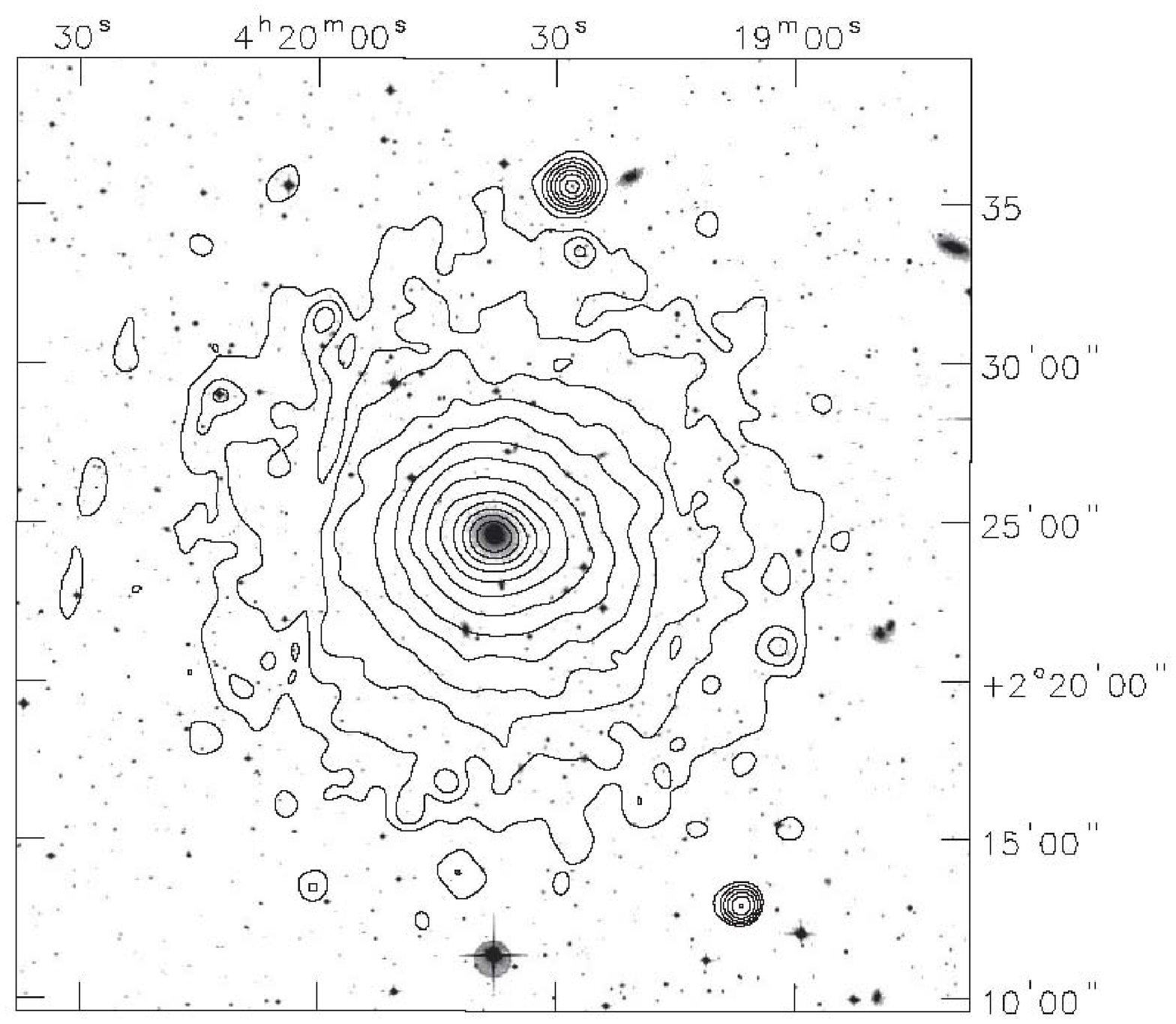}
\end{center}
\end{minipage}
 \begin{minipage}{0.5\hsize}
  \begin{center}
\vspace{0.55cm}
\includegraphics[width=0.885\textwidth,angle=0,clip]{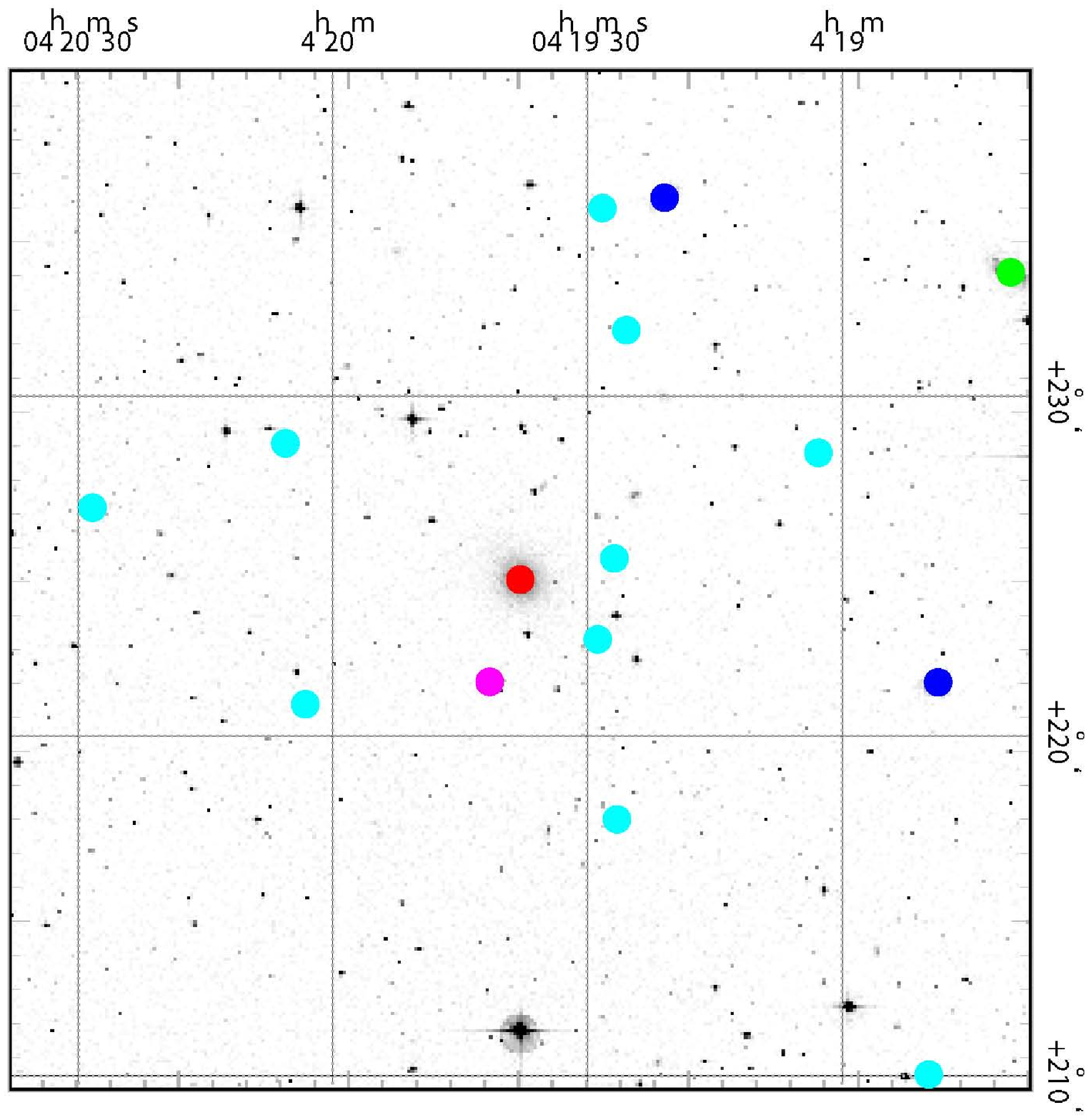}
\end{center}
\end{minipage}
\vspace{-0cm}
       %%% \FigureFile(width,height){filename}
%  \end{center}
  \caption{(left) The $0.5-4.5$ keV EPIC MOS1 surface brightness (contours) of
NGC~1550, including background, overlaid on an optical image of 
Digitazed Sky Survey (greyscale) plotted on J2000 ($\alpha$, $\delta$) 
coordinates. The X-ray image is smoothed by 
a gaussian filter with $\sigma = 20''$, but not corrected for vignetting. 
The contours are logarithmically spaced, with a step of factor 1.4.
(right) The same optical image of Digitazed Sky Survey as 
left panel, where also plotted are positions of galaxies around NGC~1550
identified in the 2MASS catalog. Colors specify their $K$-band 
magnitudes as $8<K<9$ in red, $9<K<10$ in green, $10<K<11$ in blue, 
$11<K<12$ in magenta, and $13<K$ in cyan. 
Galaxies with $12<K<13$ are not present in this sky region. 
}
\label{fig:n1550_image}
\end{figure}

%--------- 3.1 ------------------------
As shown in Figure~\ref{fig:n1550_image}~(left), 
the X-ray emission of NGC~1550 is spatially 
round, and extends at least up to a 2-dimensional radius of 
$ b \sim 14'\ (210\; $kpc$)$. 
%This suggests that NGC~1550 is a sufficiently relaxed system.
The X-ray emission centroid coincides with the position of 
the NGC~1550 nucleus within $10''\; (\sim3$ kpc).
The signal-detected region is consistent with
that obtained by {\it ASCA} (Paper I).

\begin{figure}
  \begin{center}
%\includegraphics[width=0.35\textwidth,clip]{./fig/10_0.6_80_0.5_0-600_best_rebin.epsi}
%\includegraphics[width=0.5\textwidth,clip]{./fig/n-square.ps}
    %\FigureFile(70mm,70mm){./fig/10_0.6_80_0.5_0-600_best_rebin.epsi}
    %%% \FigureFile(width,height){filename}
   \includegraphics[width=0.45\textwidth,angle=0,clip]{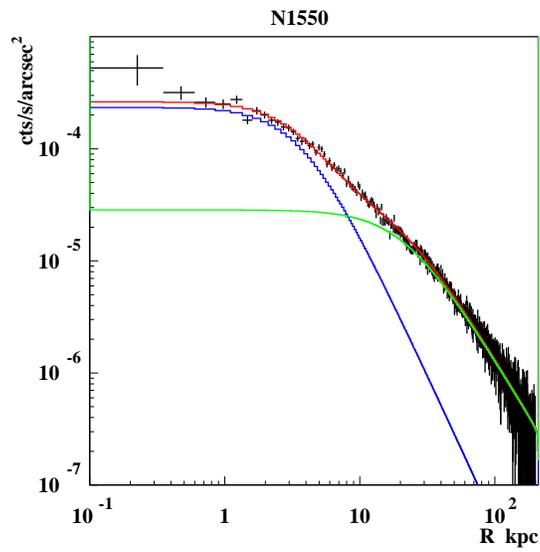} 
  \end{center}
  \caption{The $0.5-4.5$ keV background-subtracted and 
vignetting-corrected X-ray surface brightness of
NGC~1550 created with the MOS1 data (black crosses), 
shown together with the best-fit double-$\beta$ model (red). 
The data were azimuthally averaged around the NGC~1550 nucleus.
The constituent two $\beta$ components are shown in blue and green.}
\label{fig:n1550_sb}
\end{figure}

Figure~\ref{fig:n1550_sb} shows the azimuthally-averaged $0.5-4.5$ keV 
radial surface brightness profile $\Sigma(b)$, taken with MOS1 
and presented after
subtracting the blank-sky background as prepared above. 
We fitted the surface brightness profile by a $\beta$ model 
convolved with point-spread function (PSF).
The position- and energy-dependence of the PSF was quoted from the 
{\it XMM-Newton} calibration note EPIC-MCT-TN\_011 (Ghizzardi 2001).
The single-$\beta$ model, however, did not reproduce the radial profile well,
with $\chi^2/\nu=1446/599$ ($\beta = 0.45$, and the core radius of
$r_{\rm c}= 5$ kpc). Therefore, we added another $\beta$ componet 
to the model as
\begin{equation}
\Sigma(b) = \Sigma_1 \left\{ 1+\left(\frac{b}{r_1}\right)^2 \right\}^{-3\beta_1+1/2} + \Sigma_2 \left\{ 1+\left(\frac{b}{r_2}\right)^2 \right\}^{-3\beta_2+1/2}, \label{eqn:doublebeta_sb}
\end{equation}
where $\Sigma_1$ and $\Sigma_2$ are normalization constants, while
$r_1$ and $r_2$ are core radii.
This ``double-$\beta$'' model has been successful,
yielding $\chi^2/\nu=677/598$. 
The inner $\beta$ component is characterized by 
$\beta_{1} = 0.60 \pm 0.05$ and a core radius of 
$r_{1}= 3 \pm 1$ kpc, while those of the outer component are
$\beta_{2} = 0.50 \pm 0.05$ and $r_{2} = 20 \pm 3$ kpc.
These two components are shown separately in Figure~\ref{fig:n1550_sb}. 

In Paper I, a single-$\beta$ fit to the {\it ASCA} data was
successful, and yielded 
$\beta = 0.47 \pm 0.01$ and $r_{\rm c} = 16 \pm 1 $ kpc. 
The value of $\beta$ derived with {\it ASCA} is consistent with 
that of the outer $\beta$-component ($\beta_{2}$) obtained here, 
while the core radius of {\it ASCA}
falls between $r_1$ and $r_2$ of {\it XMM-Newton}.
We interpret that the angular resolution of {\it ASCA} was not
sufficient to resolve the inner-$\beta$ component, which is
seen within $\sim 15$ kpc or $\sim 1'$.

According to Sun et al. (2003), the surface brightness profile
of NGC~1550 obtained with {\it Chandra} cannot be reproduced by
a single-$\beta$ model either, and requires a double-$\beta$
modeling.
The derived outer component has $\beta = 0.48$ and 
$r_{\rm c} \sim 26$ kpc, while the inner one
is not described in their paper. Thus, the double-$\beta$ property,
found with {\it XMM-Newton}, is also supported by the {\it Chandra} 
data. Sun et al. (2003) also found an excess emission
within 1 kpc even above the double-$\beta$ fit; this is also
suggested in our result, although the angular scale of 1 kpc, or
$\sim 4''$, is comparable to the PSF of {\it XMM-Newton}.

%--------- 3.2 ------------------------
\subsection{\it Spectral analysis}
%--------- 3.2 ------------------------
In order to investigate global spectral properties of NGC~1550,
we collected photons over a circular region of $14'$
in radius from the galaxy nucleus. The background was subtracted
in the same way as described in \S\ 2, applying the same
data integration region to the background dataset as the
on-source data accumulation. 
The spectra obtained with the three EPIC detectors were 
fitted with an absorbed vMEKAL model (Kaastra \& Mewe 1993) 
or an absorbed vAPEC model (Smith et al. 2001), using XSPEC 11.3.2. 
To secure a stable fit convergence to physically reasonable 
parameter values, we fixed abundances of He, C and N to one solar, 
constrained those of other metals from zero to 3 solar, and 
fixed the column density to the Galactic value 
($1.15\times 10^{21}$ cm$^{-2}$). The employed energy range is
$0.5-5.0$ keV for MOS, and $0.7-5.0$ keV for PN. The $0.5-0.7$ keV 
energy region of PN was excluded, because of systematic residuals below 
Fe-L lines often found in PN spectra of bright clusters (Tamura et al. 2004).

We have found that a reliable estimation of oxygen abundance
requires a particular caution. This is because the oxygen 
K$_{\alpha}$ lines of NGC~1550 are weak, and the model
continuum fitted over the $0.5-5.0$ keV range (or $0.7-5.0$ keV for PN) 
often becomes slightly ($\sim 5$\%) higher (or lower) around 
the oxygen lines than the true continuum, probably due to
calibration uncertainties. In this case, 
the oxygen abundance is forced to be lower (higher) than the true 
value with little allowance, because the overall model 
normalization is tightly constrained by the entire continuum 
and hence cannot vary freely. 
In fact, the oxygen abundance and the continuum normalization 
exhibit strong and negative coupling. 
In order to avoid this problem, we estimated the oxygen abundance
by fitting the spectra over a limited energy range of
$0.5-0.8$ keV, with all parameters except the oxygen abundance
and the oxygen and continuum normalizations fixed to the best-fit 
values obtained through the full-band fitting. Although this would not be an
orthodox method to evaluate a fitting parameter, it generally
gives a more conservative and unbiased oxygen abundance. 

We firstly fitted the spectra of the three 
EPIC detectors separately with a single temperature vAPEC model, 
in order to study systematic differences among the three EPIC detectors.
The temperatures derived with PN, MOS1, and MOS2 are 
$kT = 1.26_{-0.01}^{+0.01}\; {\rm keV}$, 
$1.28_{-0.01}^{+0.01}\; {\rm keV}$, and
$1.28_{-0.01}^{+0.01}\; {\rm keV}$, respectively. 
The iron abundances from PN, MOS1, and MOS2 are
$Z_{\rm Fe} = 0.23_{-0.01}^{+0.01}\; {\rm solar}$,  
$0.26_{-0.02}^{+0.01}\; {\rm solar}$,  and
$0.25_{-0.02}^{+0.02}\; {\rm solar}$, respectively.  
Thus, the three EPIC detectors give consistent results within
their statistical errors. In the following analysis, we therefore
fit the specta of the three detectors simultaneously in order to 
improve the statistics.

\begin{figure}
  \begin{center}
%     \FigureFile(80mm,80mm){./fig/0-14_fit/n1550_0-14_sys.ps}
    \includegraphics[width=0.45\textwidth,angle=0,clip]{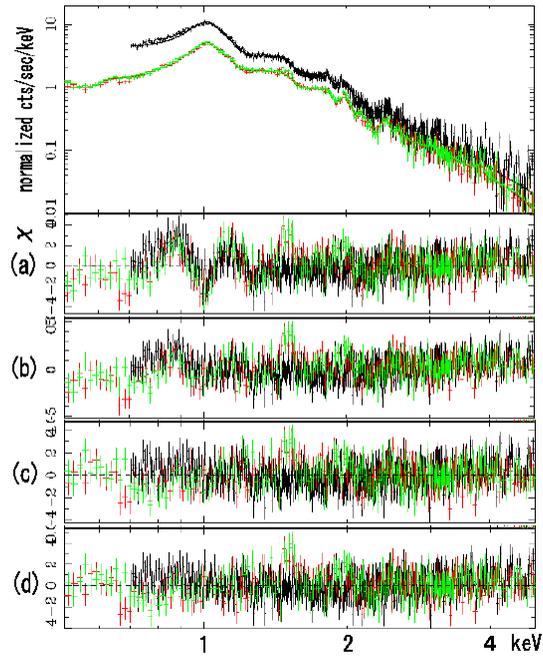}  
  \end{center}
  \caption{Response-inclusive $0.5-5.0$ keV PN (black), 
MOS1 (red) and MOS2 (green) spectra from the $0'-14'$ region, 
simultaneously fitted with; (a) an absorbed vMEKAL model,
(b) an absorbed vAPEC model, (c) a two-temperature vMEKAL model,
and (d) a two-temperature vAPEC model. 
In the two temperature fits, the metal abundances are tied between the two 
components, while the temperatures and normalizations are left free.}
\label{fig:spec-models}
\end{figure}

Figure~\ref{fig:spec-models} shows the results of the simultaneous
fit using the vAPEC and vMEKAL models, and Table~\ref{table:fit-result} 
summarizes the best-fit parameters.
Although the obtained parameters are consistent with those derived
with {\it ASCA} (Paper I), neither one-component model (vMEKAL or vAPEC) 
successfully reproduced the spectra. Since the fit residuals are seen 
around Fe-L line regions, a multi-temperature condition is suggested.
Accordingly, we introduced a two-component model, consisting of two
vAPEC (or vMEKAL) components with different free temperatures and
normalizations. The two components are assumed to share the same
metal abundances, and are subjected to the same Galactic absorption.
Then, the fit has been improved and became acceptable
(Figure 5c, 5d, and Table 1), with a $<$ 1\% probability for 
this improvement to be accidental.
We hereafter adopt the vAPEC code, because it can generally reproduce
the Fe-L lines better (e.g. Matsushita et al. 2002); our results
remain essentially the same even if we use the vMEKAL model.
With the {\it ASCA} data, we derived the metal abundance to be $0.3 \pm 0.1$ 
using the MEKAL model (Paper I). This is consistent with the iron and 
silicon abundances in Table~\ref{table:fit-result}, although the 
{\it ASCA} results are subject to larger errors.

Although we constrained the two vAPEC components to have the same 
metal abundances, this may not be necessarily warranted. Therefore,
we tentatively relaxed this constraint, and found that 
the iron abundances of the hot and cool components became 
$Z_{\rm Fe,High} = 0.34 \pm 0.02$ and $Z_{f\rm Fe,Low} = 0.43 \pm 0.05$, 
respectively. Therefore, the tied value of $Z_{\rm Fe} = 0.34 \pm 0.01$ 
is considered to be appropriate for the hot component, while it may 
deviate from the true iron abundance of the cool component by $-20$\%. 
Although this difference exceeds statistical errors, the fit is not 
improved significantly,  because the F-test indicates a 88\% 
probability of the difference by chance. Furthermore, 
the oxygen abundances became poorly constrained as 
$Z_{\rm O,High} = 0.18 \pm 0.07$ and $Z_{\rm O,Low} = 0.71 \pm 0.68$.
Considering these, we retain our original assumption that the
two vAPEC components share the same metal abundances. 
Even when taking into account the possible $\sim 20$\% systematic
error in the metal abundances of the cool component, our comparison
between the metal distributions with that of the galaxy light,
described in \S 5, remains essentially unaffected.

\begin{figure}
 \begin{minipage}{0.5\hsize}
  \begin{center}
\includegraphics[width=0.7\textwidth,angle=90,clip]{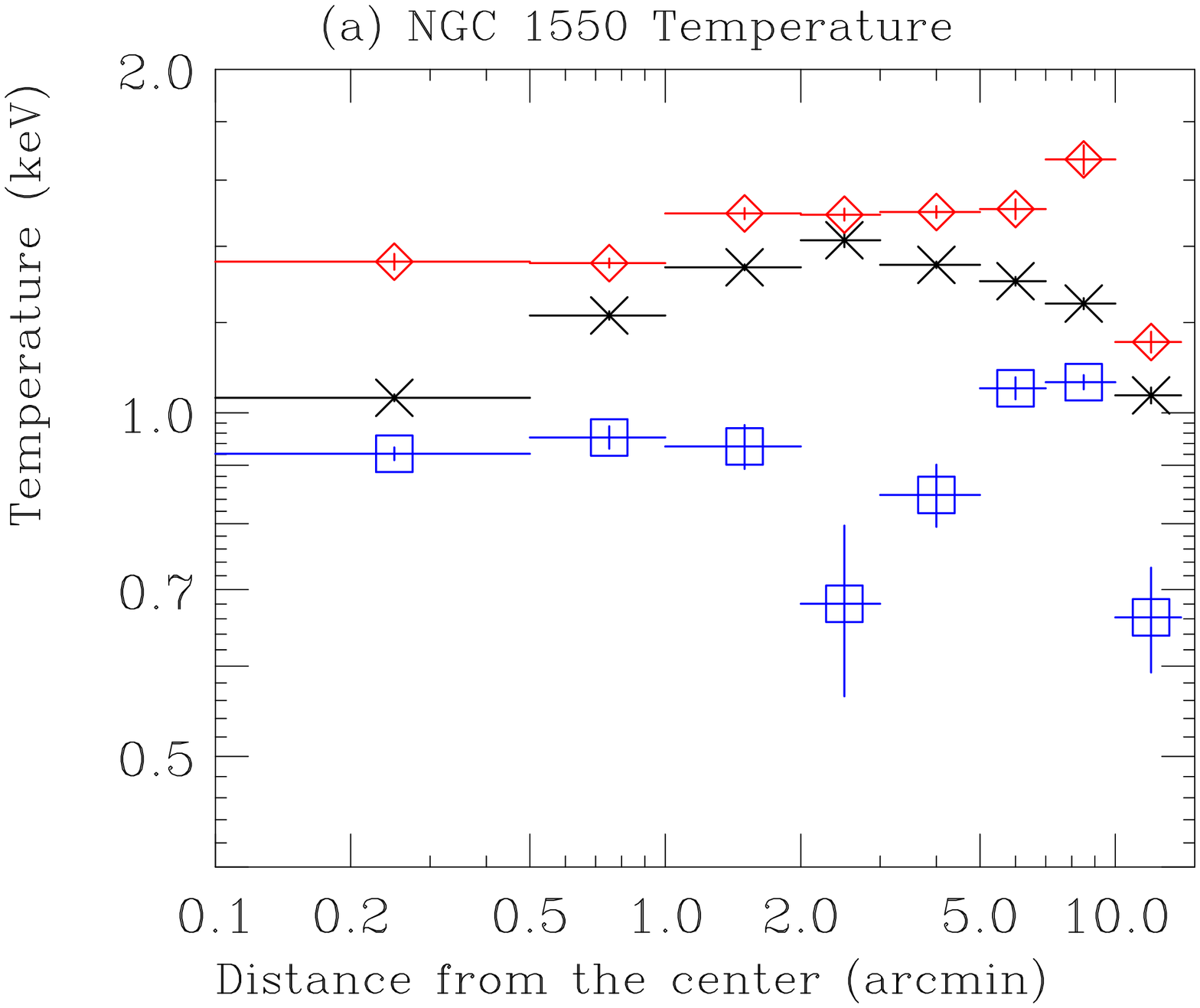}
 \end{center}
 \end{minipage}
  \begin{minipage}{0.5\hsize}
  \begin{center}
\includegraphics[width=0.7\textwidth,angle=90,clip]{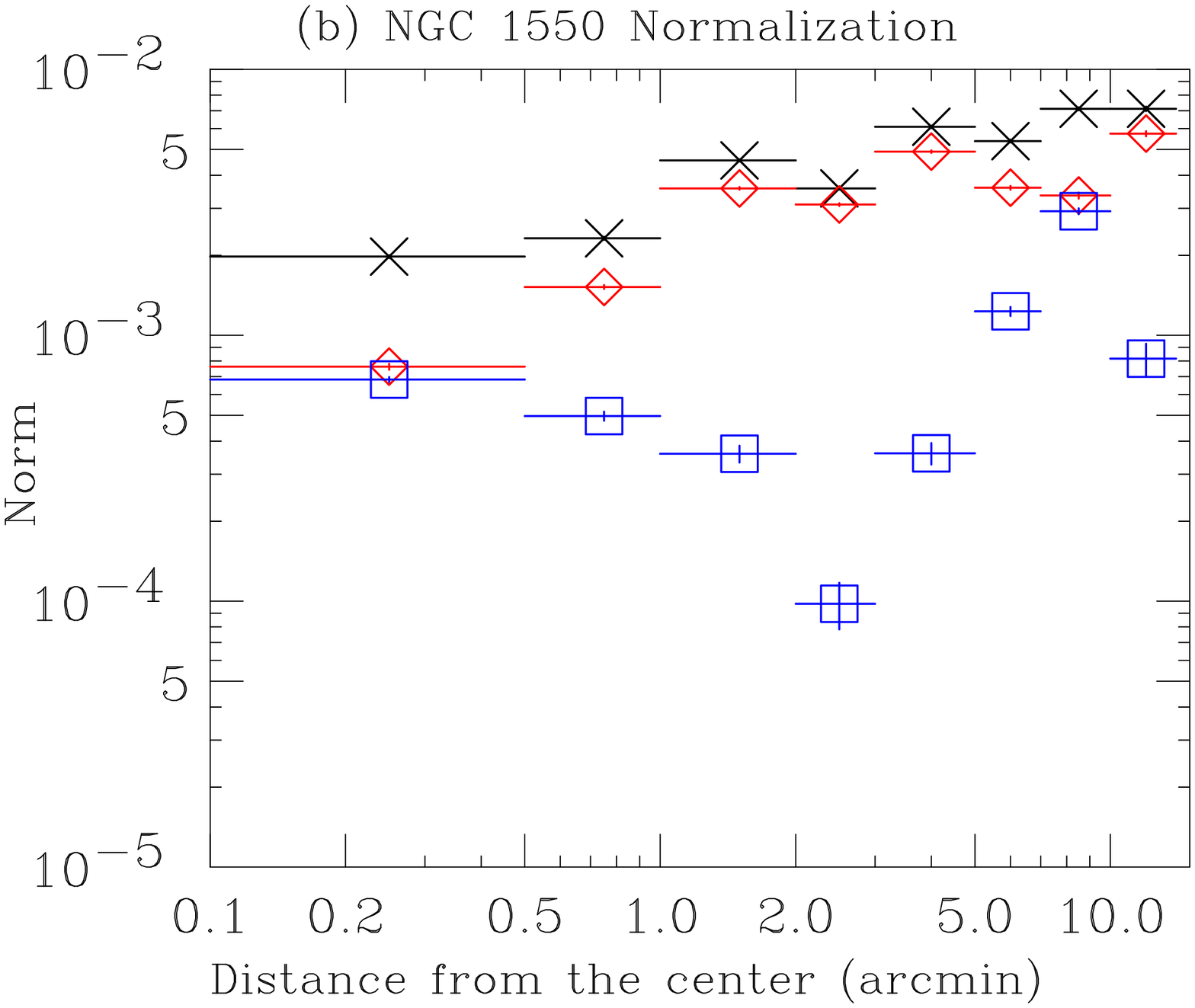}
 \end{center}
 \end{minipage}
\begin{minipage}{0.5\hsize}
\vspace{0.5cm}
  \begin{center}
\includegraphics[width=0.7\textwidth,angle=90,clip]{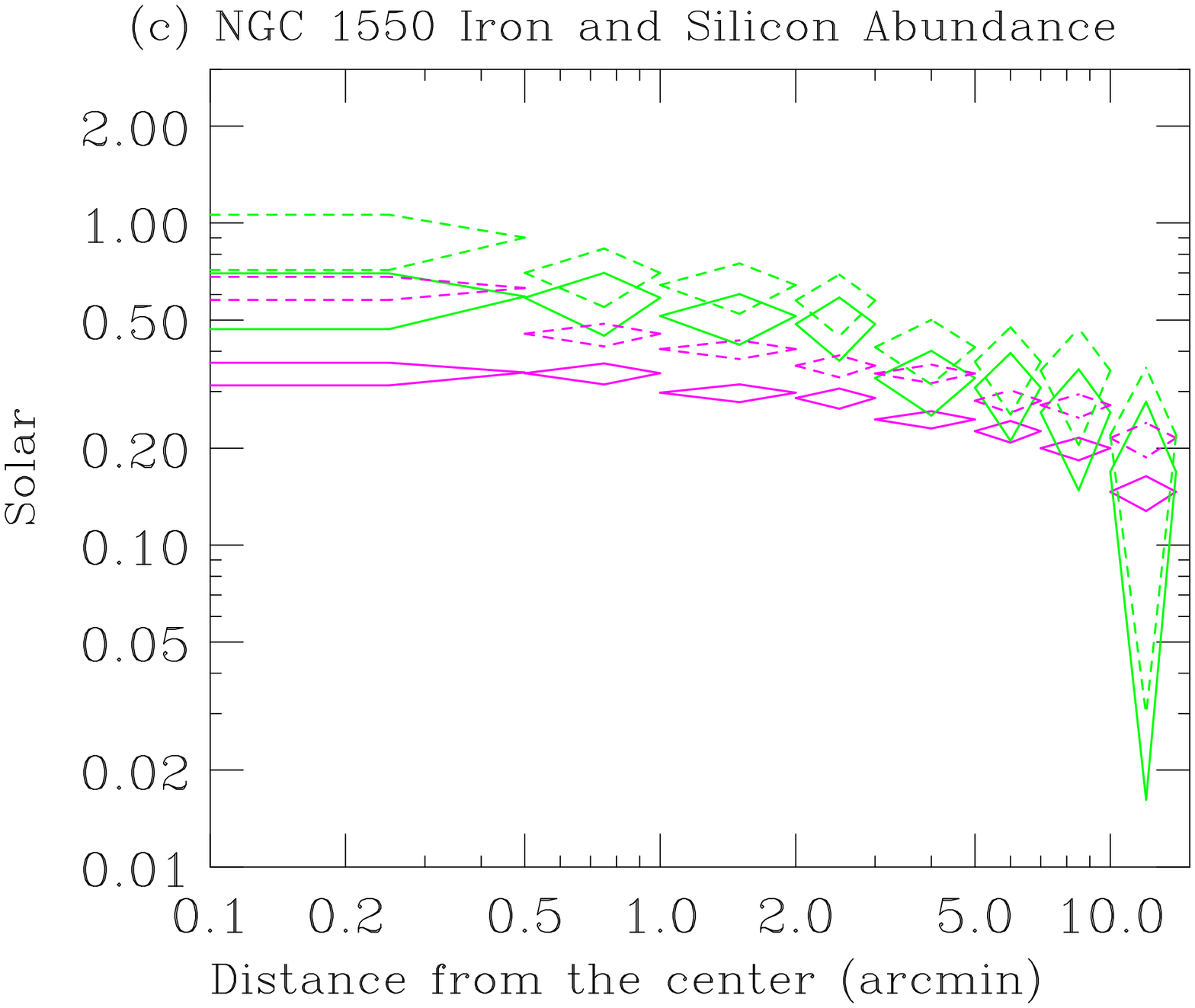}
 \end{center}
 \end{minipage}
 \begin{minipage}{0.5\hsize}
\vspace{0.5cm}
  \begin{center}
\includegraphics[width=0.7\textwidth,angle=90,clip]{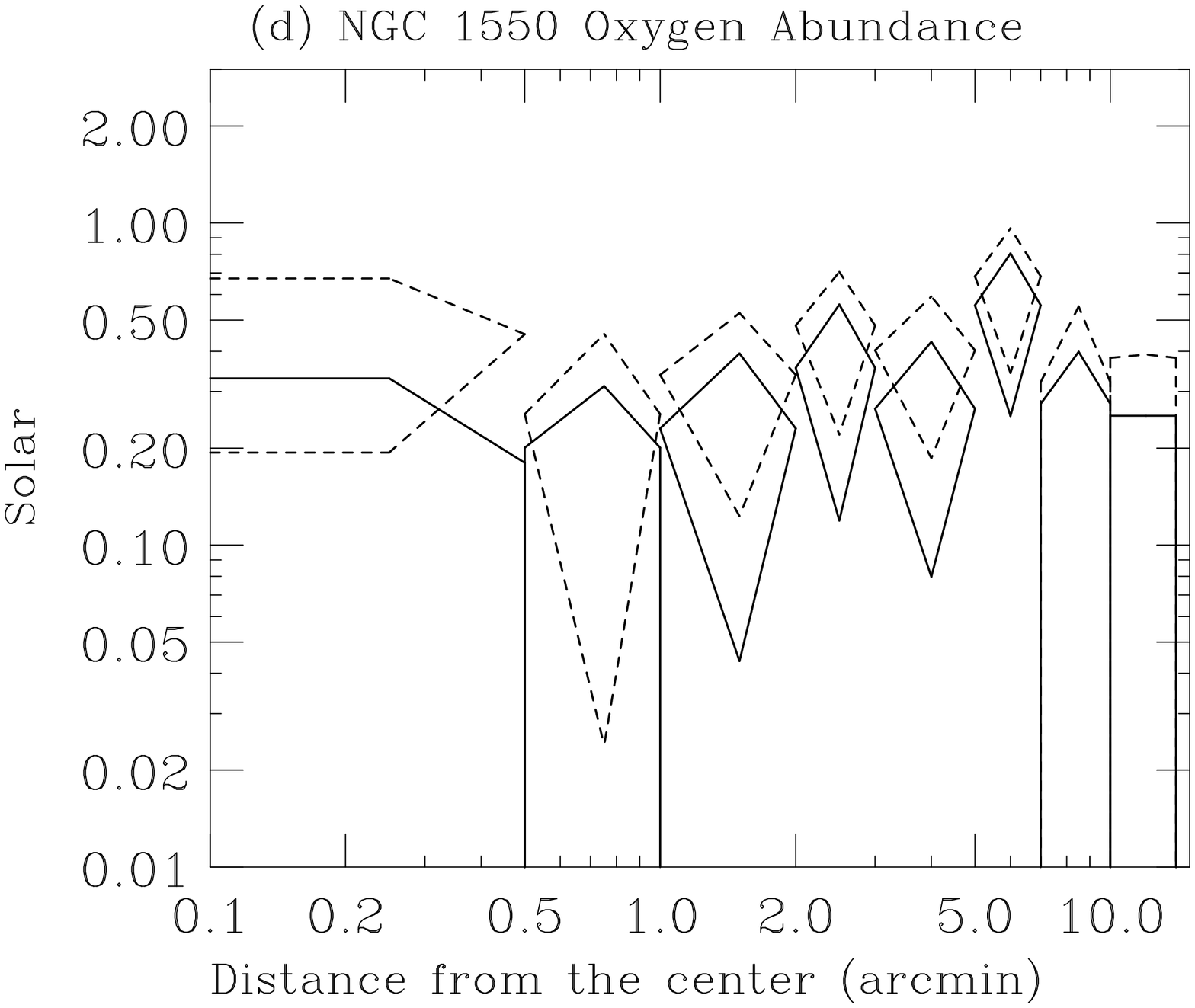}
 \end{center}
 \end{minipage}
 \begin{minipage}{0.5\hsize}
\vspace{0.5cm}
  \begin{center}
\includegraphics[width=0.7\textwidth,angle=90,clip]{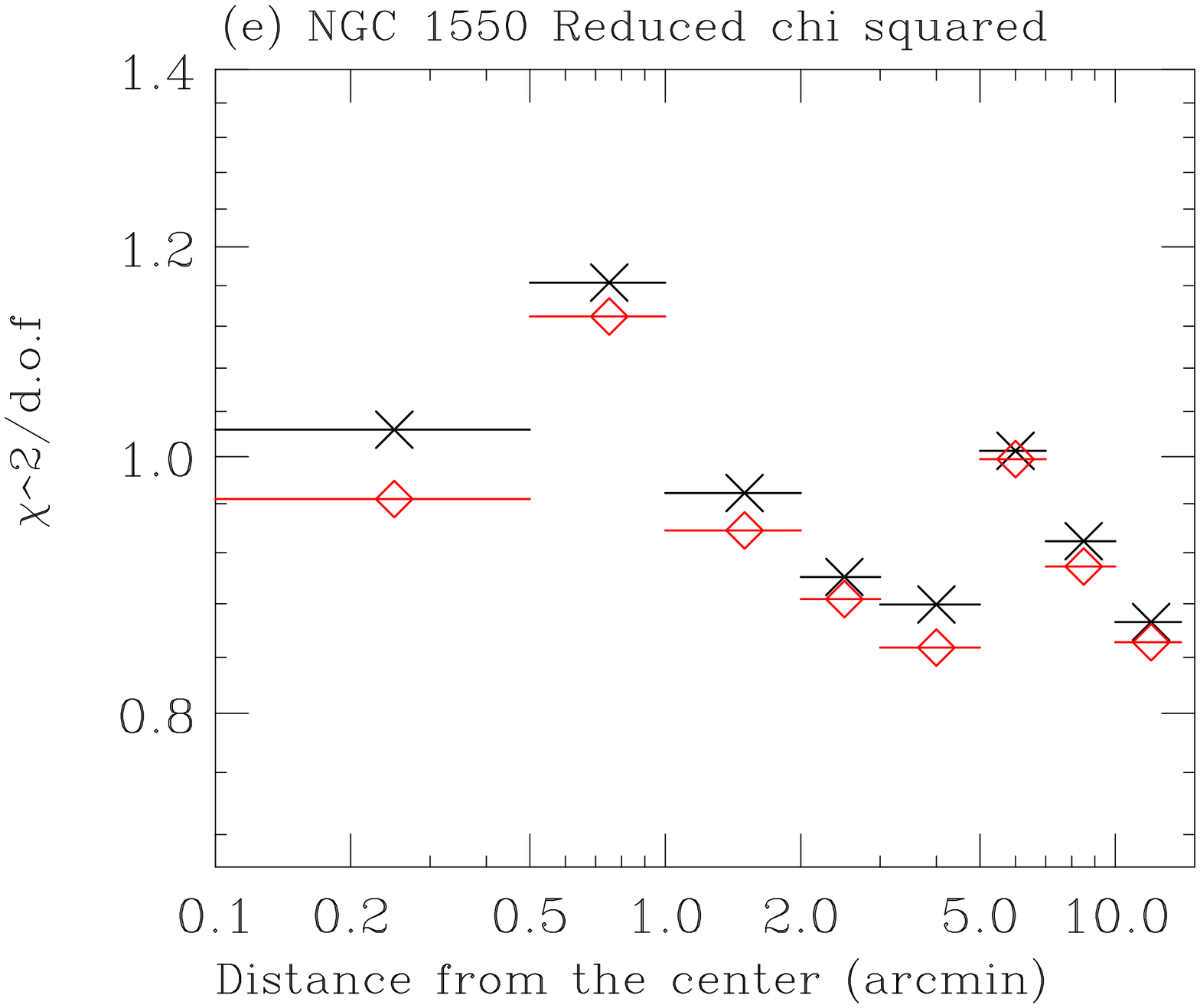}
 \end{center}
 \end{minipage}

  \caption{Radial profiles of the X-ray spectroscopic parameters,
derived with 8 annular spectra.
(a) The temperature derived with the single (black) and
double (red and blue) vAPEC fits.
(b) The vAPEC normalizations (proportional to emission measure), 
with the same color definitions as in panel (a).
(c) The iron (magenta) and silicon (green) abundances obtained from
the single (solid diamonds) and double (dashed diamonds) vAPEC fits.
(d) The oxygen abundances, with the same definitions of solid and dashed
diamonds as (c). 
(e) The reduced chi-squared from the single (black) and
double (red) vAPEC fits. 
}
\label{fig:annular-prof}
\end{figure}

As a next step, we studied spatial variations of the temperature 
and metal abundances. For this purpose, we divided the image into 
eight concentric annular regions, centered on the X-ray emission peak 
(and hence on the central galaxy). 
The regions were chosen so that each of them contains more than 5,000 photons 
after the background subtraction. Then, the spectra of the eight regions 
were fitted in the same way as for the spatially integrated spectra.  
Figure~\ref{fig:annular-prof} shows radial profiles of the temperature, 
abundances and the reduced chi-squared, derived by fitting 
the individual annular spectra (simultaneous among the three detectors) 
with one-temperature and two-temperature vAPEC models. 
The temperature determined with the one-component model increases
mildly from 1.0 keV at the center to 1.4 keV at $\sim 3'$ (45 kpc),
and then returns to 1.0 keV at $\sim 10'$ (150 kpc).  
Though somewhat depending on the modeling (single vs. double temperatures),
the oxygen abundance is low and spatially rather constant, while those of 
the other metals clearly increase toward the center by a factor
of 3 to 5. The iron and silicon profiles
are rather similar, regardless of the temperature modeling,
although the double-temperature fit yields systematically 
higher abundances than the single-temperature one.
In all regions except for the outermost one, the two temperature
model gives a better reduced chi-squared, and the $F$-test 
indicates that the probability of these improvements caused by chance 
is less than 1\% in six regions except the $5.'0-7.'0$ and $10.'0-14.'0$ ones. 
The two temperatures (1.5 and 0.95 keV; Table~\ref{table:fit-result})
found with the region-integrated spectra can be understood reasonably
as weighted means of the two temperatures specified by the eight
annular spectra.

Although the EPIC data thus prefers the two-temperature modeling
to the single-phase views, this could simply be due to projection effects.
We accordingly deprojected the eight annular spectra into eight shell 
spectra using  ``onion-peeling'' method (e.g. Buote et al. 2003), 
in which the contribution from the outer shell is subtracted 
progressively toward smaller radii.
The emission from the region lying outside the outermost radius ($14'$)
was estimated by assuming that the spectrum of the outermost annulus
extends with its surface brightness following the outer $\beta$ component
derived in \S 3.1.
In the model fitting to the deprojected spectra, 
we set initial parameter values to the best-fit ones obtained with the
annular spectra, and again fixed or constrained the metal abundances 
in the same manner as the annular analysis. 
The obtained profiles of the model parameters,  presented 
in Figure~\ref{fig:shell-prof}, 
are similar to those obtained from the annular spectra.
The two modelings (single vs. two temperatures) are both acceptable
essentially in all shells, and their fit goodness no longer shows
the significant difference. 
This is due partially to the removal of the projection effects, and
partially to reduced data statistics.
Therefore, the plasma in each shell can be represented by a 
single-temperature collisionally-ionized plasma model, although local 
two-temperature (or multi-temperature) conditions are not necessarily 
ruled out.

%%%4. Near Infra-Red analysis %%%%%
\section{Near Infra-Red Data Analysis}
%%%%%%%%%%%%%%%%%%%%%%%%%%%%%%%%%%%
In order to estimate the galaxy light profile around NGC~1500,
we utilezed the 2MASS (Two Micron All Sky Survey) data, 
which are accessible via NASA/IPAC Exgragalactic Database.
In the 2MASS catalog, 33 galaxies (with $K$-band magnitude
less than 14) have been found to be lying within $34'$ of the NGC~1550 
nucleus; at the distance of NGC~1550, this projected radius 
corresponds to 500 kpc. The sky positions of the galaxies 
in $30'\times 30'$ region around NGC~1550 are plotted in 
Figure~\ref{fig:n1550_image}~(right).
Among the 33 galaxies, 7 galaxies listed in Table~\ref{table:member} 
are no fainter in $K$-band than NGC~1550 by 2.5 magnitude.
These galaxies, except MCG+00-11-050 (of which the radial velocity 
is unknown), have radial velocities within $\pm 500$ km s$^{-1}$ of 
that of NGC~1550, and  are likely to be members of the 
``NGC~1550 group''. Three galaxies (IC~0366, UGC~03011, and MCG+00-11-050) 
of the 7 are within the X-ray signal-detected region ($14'$ in radius).

Figure~\ref{fig:k-lumi_prof} (red diamonds) shows the
2-dimensional integrated $K$-band profile around NGC~1550,
constructed from NGC~1550 itself and the 33 galaxies. 
In this figure, the surface brightness profile of NGC~1550 
is taken into account by using a public 2MASS 
background-subtracted $K$-band image of NGC~1550, 
while the other galaxies were treated as point sources. 
Since this profile must be contributed by background objects,
we estimated the number of background galaxies expected 
in the $34'$-radius region, based on the number counts of galaxies 
with $K \leq 14$ from the entire 2MASS sample (Bell et al. 2003).
The $K$-band magnitude was converted to luminosity assuming the 
absolute $K$-band solar magnitude of 3.39 (Kochanek et al. 2001)
and the same distance to all the galaxies as that to NGC~1550.
Figure~\ref{fig:k-lumi_prof}~(blue boxes) is the estimated
two-dimensionally integrated background $K$-band luminosity profile. 
Even at the angular distance of $34'$, the background is about 
5 times lower than the integrated $K$-band luminosity of 
the NGC~1550 system. Within $34'$ (510 kpc) and $14'$ (210 kpc), 
the central galaxy carries $39$\% and 70\%
of the background-subtracted $K$-band luminosity, respectively. 

%%%5. Mass profiles %%%%%
\section{Spatial Distributions of Mass Components}
%%%%%%%%%%%%%%%%%%%%%%%%%%%%%%%%%%%
%--------- 5.1 ------------------------
\subsection{\it Mass profiles}
%--------- 5.1 ------------------------
From the parameters of the X-ray emitting gas obtained in the spatial and
spectral analyses (\S 3.1 and \S 3.2),
we derived radially integrated mass profiles of three components constituting
the NGC~1550 system; the ICM mass, the total mass, and the gaseous metal mass;
the results are shown in Figure~\ref{fig:mass-prof}a.
Below, we describe how these results have been obtained.

As a step to calculate the ICM mass profile, we started
from the best-fit double-$\beta$ model of equation~\ref{eqn:doublebeta_sb}. 
We analytically deprojected it into a 3-dimensional emissivity profile as
\begin{equation}
\epsilon(r) = \epsilon_1 \left\{ 1+\left( \frac{r}{r_1}  \right)^2   \right\}^{-3\beta_1} + \epsilon_2 \left\{ 1+\left( \frac{r}{r_2} \right)^2   \right\}^{-3\beta_2}, \label{eqn:doublebeta_emissivity}  
\end{equation}
where $r$ is 3-dimensional radius from the emission centroid, and
$\epsilon_i$ ($i=1,2$) are constants which are
determined uniquely by $\Sigma_i$, $r_i$, and $\beta_i$ of 
equation~(\ref{eqn:doublebeta_sb}).
Then, the emissivity is rewritten as $\epsilon = \Lambda(T,Z) n_{\rm e}^2$, 
where $\Lambda(T,Z)$ is the cooling function specified by the 
measured temperature and metallicity, and $n_{\rm e}$ is the electron
density. Using this $n_{\rm e}$, the ICM mass density profile is written as 
\begin{equation}
\rho_{\rm gas}(r) = \mu m_{\rm p} n_{\rm e}(r) =  \mu m_{\rm p} \sqrt{\frac{\epsilon(r)}{\Lambda(T,Z)}}\ \ . \label{eqn:gas-density}
\end{equation}
Here $\mu = 1.2$ is the mean molecular weight
(i.e., the mean number of nucleon per electron). 
Values of $\Lambda(T,Z)$ were estimated for each shell region,
utilizing $T$ and $Z$ measured in \S\ 3.2, and referring 
to tables in Sutherland \& Dopita (1993). 
Finally, we obtained the radially integrated ICM mass of Figure~\ref{fig:mass-prof}a, 
in the standard way, as
\begin{equation}
M_{\rm gas}(r) = \int_0^r \rho_{\rm gas} \ 4 \pi r'^2 dr'\ \ . \label{eqn:gas-mass}
\end{equation}
In Figure~\ref{fig:mass-prof}a, the gas mass profile outside $14'$
was calculated assuming that the emissivity profile follows equation~(\ref{eqn:doublebeta_emissivity}) with the same temperature and metal abundance as 
those in the $10'$--$14'$ region. 
In paper I, the ICM mass inside $r = 220$ kpc was measured to be 
$3.4 \times 10^{11} {\rm M}_{\odot}$; at the same radius, we now
obtain a value as $M_{\rm gas} = 7.6 \times 10^{11} {\rm M}_{\odot}$.
The difference is due to the inner $\beta$ component, which is
apparent in Figure~\ref{fig:n1550_sb} and included in the 
{\it XMM-Newton} analysis. The ICM mass derived here is
consistent with that from {\it Chandra} (Sun et al. 2003).

The metal mass density profiles in the ICM were derived by multiplying 
$\rho_{\rm gas}(r)$ of equation~(\ref{eqn:gas-density}) with the measured 
metal abundances. 
For example, the radial profiles of the iron mass density, $\rho_{\rm Fe}(r)$,
can be derived by multiplying the ICM mass profile with 
the iron abundance profile as
\begin{eqnarray}
\rho_{\rm Fe}(r) = A_{\rm Fe} m_{\rm p} n_{\rm Fe}(r)  
              = A_{\rm Fe}\left( \frac{n_{\rm Fe}}{n_{\rm H}}\right)_{\odot}\; Z_{\rm Fe}(r)\; f_{\rm H}\; \rho_{\rm gas}(r), \label{eqn:iron-density}
\end{eqnarray}
where $A_{\rm Fe}$ is the averaged atomic weight of Fe, 
$(n_{\rm Fe}/n_{\rm H})_{\odot} = 4.68 \times 10^{-5}$ is 
the number ratio of iron to hydrogen under one solar abundance 
(Anders \& Grevesse 1989), $Z_{\rm Fe}(r)$ is the measured iron abundance,
and $f_{\rm H} = 0.75$ is the hydrogen mass fraction in the ICM.
If another definition of solar abundance such as Grevesse \& Sauval (1998)
is adopted, $Z_{\rm Fe}(r)$ can change, but $\rho_{\rm Fe}(r)$ remains the
same, because the change in $Z_{\rm Fe}(r)$ is canceled out by that in
$\left( \frac{n_{\rm Fe}}{n_{\rm p}}\right)_{\odot}$; that is,  $\rho_{\rm Fe}(r)$
is independent of the definition of the ``solar abundance''.
As to $Z_{\rm Fe}(r)$ in equation~(\ref{eqn:iron-density}), we used the
results from the single temperature fits to the deprojected 
shell spectra. 
However, as can be inferred from Figure~\ref{fig:shell-prof}c,
the results do not change by more than 60\% even when the 
two-temperature fits are employed. 

At $r = 210$ kpc, the integrated 
masses of iron, silicon, and oxygen imply $0.18^{+0.04}_{-0.05}$, 
$0.27^{+0.20}_{-0.13}$, and $0.29^{+0.54}_{-0.25}$ solar abundances, 
respectively, when divided by the integrated gas mass. 
In Figure~\ref{fig:mass-prof}a, iron and silicon exhibit very
similar integrated mass profiles, reflecting similar abundance 
profiles (Figure~\ref{fig:shell-prof}c). In contrast, the integrated oxygen
mass increases somewhat more steeply in Figure~\ref{fig:mass-prof}a
than those of iron and silicon. Obviously, this is because the
oxygen abundance is radially rather constant, unlike iron and silicon
which show the central abundance enhancements. In a word, the oxygen
mass profile is very similar to that of the ICM.

The integrated total mass profile was obtained assuming hydrostatic pressure 
equilibrium of the ICM in the gravitational potential, 
using the formula as
\begin{equation}
M_{\rm tot}(r)=-\frac{kTr}{\mu Gm}
             \left( \frac{d \ln{\rho_{\rm gas}}}{d \ln{r}}+\frac{d \ln{T}}{d \ln{r}}\right) \ \ , \label{eqn:tot-mass}
\end{equation}  
where $k$ is the Boltzmann constant and $G$ is the constant of gravity.
In this calculation, the temperature $T$ refers to the values measured with
the deconvolved spectra. We evaluated the term $\frac{d \ln{T}}{d \ln{r}}$ 
for each region by connecting two adjacent temperature points 
with a power-law, and substituting $\frac{d \ln{T}}{d \ln{r}}$ with 
the power-law slope averaged on both sides of each point. 
In the present case, 
$- \frac{d\; {\rm ln}\; \rho_{\rm gas} }{d\; {\rm ln}\; r}$ is
in the range of  $1.2{\bf -}2.1$, whereas 
$\frac{d\; {\rm ln}\; T }{d\; {\rm ln}\; r}$ is
typically $\lesssim 0.1$ (0.5 at maximum). 
Therefore, the proper inclusion of the 
$\frac{d\; {\rm ln}\; T }{d\; {\rm ln}\; r}$ term
reduces the estimated $M_{\rm tot}$ up to several tens percent,
in those regions where $T(r)$ is increasing outward. 
Outside $14'$, the temperature was assumed to be radially constant
and the same as that in the $10'$--$14'$ region.
At $r \sim 210$ kpc, we thus obtain the total mass as 
$2.0 \times 10^{13} M_{\odot}$, which agrees with the value
of $2 \times 10^{13} M_{\odot}$ typically found among
galaxy groups (Mulchaey et al. 1996). Also, the present result
agrees with that in Paper I; $1.6 \times 10^{13} M_{\odot}$ at
$r = 210$ kpc. 
The total mass profile obtained with {\it Chandra} (Sun et al. 2003) 
is consistent with our profile (Figure~\ref{fig:mass-prof}a)
within respective errors. 
Furthermore, the gas-mass fraction at $\sim 400$ kpc
(half the virial radius), $M_{\rm gas}/M_{\rm tot} = 0.06$, is 
consistent with those observed from galaxy groups at a 
radius of overdensity $\sim 500$ (Vikhlinin et al. 2006). 

Based on {\it ASCA} observation, Dupke \& Bregman (2005) reported
the presence of a velocity difference by $\ge 1.5 \times 10^3$ km s$^{-1}$
in the ICM of NGC~1550. If so, equation~(\ref{eqn:tot-mass}), which
assumes hydrostatic confinement of the ICM, would become no longer
valid. 
However, a claimed velocity difference of
$2400 \pm 1000$ km $s^{-1}$
in the Centaurus cluster, by the same authors 
(Dupke \& Bregman 2006), 
was not found significantly by the latest Suzaku measurements, 
with an upper limit velocity of $1400$ km s$^{-1}$ at 90\% confidence
range (Ota et al. 2007). Therefore, the reality of the reported bulk 
motion in the ICM of NGC~1550 is open, and hence we retain our original 
assumption of hydrostatic equilibrium. Even if the bulk motion is real, 
the total mass estimate would change only by a factor of $\sim 2$ 
(Ota et al. 2007), which does not affect our main results.

The present study requires a quantitative comparison of the X-ray
derived 3-dimensional radial mass profiles against visible-light
distribution representing the stellar mass profile. However, 
the galaxy distribution obtained in \S 4 is a
2-dimensional quantity expressed on the celestial sphere.
Therefore,  we have to deproject the galaxy profile into three dimensions, 
or alternatively, project the other three-dimensional 
profiles (the total mass, gas mass, and metal masses) onto two dimensions.
We have adopted the latter method, because the galaxies aroung NGC~1550
are too sparsely distributed to allow reliable deprojection.
Figure~\ref{fig:mass-prof}b shows the projected mass profiles 
in the integrated form, including the galaxy mass profile. 
In the derivation of the galalxy masses, we assumed $K$-band stellar
mass-to-light ratio ($M/L$) of all the galaxy to be 1.0 for simplicity,
altough $M/L$ in $K$-band varies approximately between 0.5 and 1.0 
according to the color of galaxies (Bell et al. 2003). 
Thus, the galaxy mass profile presented here is essentially the same as
the $K$-band luminosity profile.

As can be seen in Figure~\ref{fig:mass-prof}b, the radially-integrated
$K$-band mass-to-light ratio of the system reaches 
$75 \pm 3\; M_{\odot}/L_{\odot}$ at 210 kpc.
Considering a typical color of galaxies, this value is consistent
with the $B$-band mass-to-light ratio of $\sim 380\; M_{\odot}/L_{\odot}$
($r < 210$ kpc) derived in Paper I. These values are comparable to those
of clusters of galaxies. In fact, the mass-to-light ratio of 
galaxy clusters is typically measured to be 
$\sim 250\; M_{\odot}/L_{\odot} $ in $B$-band 
(Bartelmann \& White 2002) and $\sim 80\; M_{\odot}/L_{\odot}$ 
in $K$-band (Kochanek et al. 2003).

%--------- 5.2 ------------------------
\subsection{\it Iron-mass-to-light ratio profile}
%--------- 5.2 ------------------------  
As shown by Figure~\ref{fig:mass-prof}b, the ICM distribution
is slightly more extended than that of the total mass
(i.e. the gas fraction increasing toward the cluster outskirts), 
in agreement with previous reports (e.g. Markevitch \& Vikhlinin 1997). 
Furthermore, the figure clearly reveals that the integrated stellar luminosity 
profile is significantly flatter than the other profiles: the stellar component
is much more concentrated than the other components. 
The difference between the stellar luminosity profile and 
those of the metal masses is particularly interesting, because 
the metals should have originated from stars. 
In order to see the difference more clearly, we divided the 
integrated iron mass profile by integrated 
$K$-band luminosity profile, to obtain
so-called iron-mass-to-light ratio (IMLR) profile (Finoguenov 2000). 
The radially-integrated IMLR profile of NGC~1550, 
shown in Figure~\ref{fig:imlr-ddp}a,
drops toward the center by about two orders of magnitudes, while flattens
outside $\sim 100$ kpc. This behavior is in agreement with the
$ASCA$ result on the Centaurus cluster and Abell~1060 
(Makishima et al. 2001), suggesting that the IMLR decreases
towards the center are common among regular clusters regardless 
of their richness. The silicon-mass-to-light ratio (SiMLR) 
and oxygen-mass-to-light ratio (OMLR) profiles also decrease toward 
the center to the same degree as the IMLR.

%--------- 5.3 ------------------------
\subsection{\it Density-density plot
\label{subsec:ddp}
}
%--------- 5.3 ------------------------
The IMLR and similar quantities compare two relevant mass components
in the form of ratios between their radially-integrated profiles.
This approach is useful to grasp large-scale behavior of these
components, but may not be suitable to investigate more localized features.
For this purpose, we devised a new plot which we named 
density-density plot (DDP). A DDP plots densities of a component 
in individual shells against those of another component in the
corresponding shells. Although the spatial coordinate is lost in a DDP, 
we can directly see the relation between any two mass components in 
their differential forms.

Figure~\ref{fig:imlr-ddp}b shows DDPs between the total mass
and iron mass (red diamonds), and between the gas mass and iron 
mass (blue squares). The latter DDP indicates that the iron mass
density increases (toward the center) somewhat more steeply
than the ICM mass density, just reflecting the iron abundance
increase in Figure~\ref{fig:shell-prof}c from $\sim 0.15$ solar
at $r \sim 200$ kpc to $\sim 0.5$ solar at the center.

Interestingly, we instead find in Figure~\ref{fig:imlr-ddp}b 
a very good proportionality between the iron mass density and 
the total gravitating mass density (red diamonds) except at
the very center. In fact, the least-square fit with a power law function 
yields the best relation between these two quantities,
excluding the innermost data point, as 
\begin{equation}
\rho_{\rm Fe} = (3.2 \pm 1.6) \times 10^{-5}\; \rho_{\rm tot}^{0.95\pm 0.03}, \label{eqn:ddp-fe-tot}  
\end{equation}  
while the relation between the iron mass density and 
the ICM mass density becomes
\begin{equation}
\rho_{\rm Fe} = (4.7 \pm 3.3) \times 10^{-5}\; \rho_{\rm ICM}^{1.24\pm 0.09}. \label{eqn:ddp-fe-icm}  
\end{equation} 
In other words, the iron in the ICM traces the total mass distribution 
(except the innermost region) rather than that of the ICM density. 
In the innermost region ($r < 7.5$ kpc), 
the iron mass fraction decreases significantly relative to the total mass. 
This is likely to be related to the dominance of the stellar mass
of NGC~1550 in this region, although it is not immediately clear, 
at this stage, why the iron mass increases less prominently than 
the stellar mass.

Incidentally, equation~(\ref{eqn:ddp-fe-tot}) and  
equation~(\ref{eqn:ddp-fe-icm}) can be combined into a relation
between the ICM mass density and the total mass density, as 
$\rho_{\rm ICM} \propto \rho_{\rm tot}^{0.77 \pm 0.08}$. This is reasonable,
because the index of 0.77 is close to the typical value of 
$\beta \sim 0.7$ (Mohr et al. 1999) found among clusters, where $\beta$
is the same index as employed in equation~(\ref{eqn:doublebeta_sb}).

In order to investigate the above issue at the innermost region, let
us compare densities between the galaxy $K$-band luminosity 
and the iron mass. For the same reason as described in \S5.1,
we use the 2-dimensional form, which we named Surface Density-Surface 
Density Plot (SDSDP): a SDSDP plots the surface density of a 
component in the individual 2-dimensional annular regions against 
those of another component.
Figure~\ref{fig:imlr-ddp}c shows a SDSDP between the
galaxy $K$-band luminosity ($\Sigma_{\rm gal}$) 
and the iron mass ($\Sigma_{\rm Fe}$). Thus, the
innermost two regions having $> 10^8 L_{\odot}$ kpc$^{-2}$
are more optically luminous than the other regions by
$\sim 2$ orders of magnitude. This demonstrates the
characteristic of the NGC~1550 system as an IXEG. 
In contrast, the iron surface density does not exhibit such a
steep increase to the center, so that $\Sigma_{\rm Fe}$ is
not proportional to $\Sigma_{\rm gal}$ particularly toward
the center. This is another expression of the IMLR profile 
(Figure~\ref{fig:imlr-ddp}a): in the central region of the NGC~1550
system, iron in the ICM becomes much depleted relative to the 
stellar light, even though it moderately increases relative to the
ICM mass density.

As can be readily inferred, the relation between silicon and the stellar mass 
is essentially the same as that between iron and the stars.
Oxygen (though not shown) is even more depleted in the
central region than iron and silicon, because it lacks the
central abundance increase.

%%%6. Discussion %%%%%%%%%%%%%%%%%%
\section{Discussion}
%%%%%%%%%%%%%%%%%%%%%%%%%%%%%%%%%%%

\begin{figure}
  \begin{minipage}{0.5\hsize}
  \begin{center}
\includegraphics[width=0.7\textwidth,angle=90,clip]{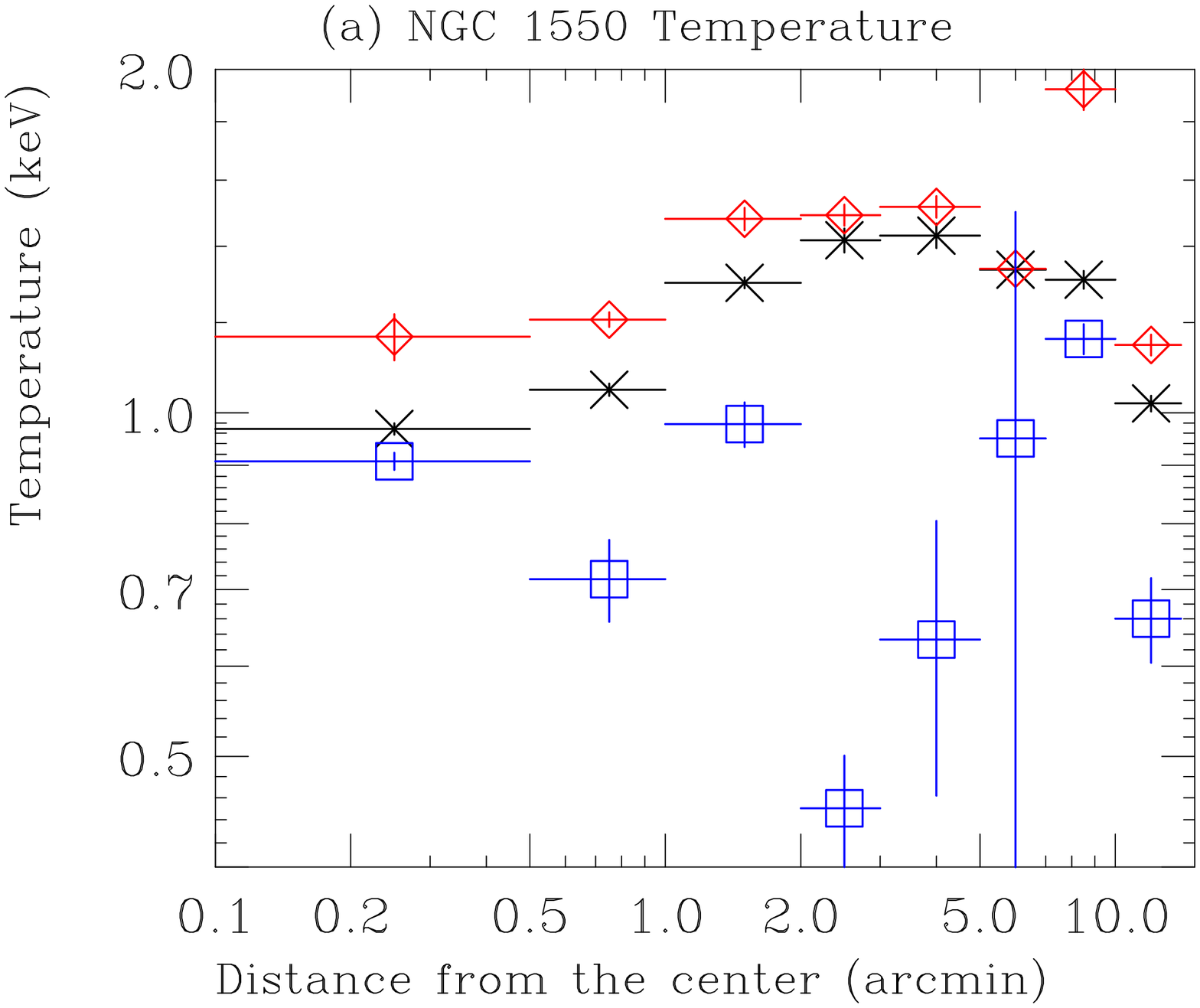}
  \end{center}
  \end{minipage}
  \begin{minipage}{0.5\hsize}
  \begin{center}
\includegraphics[width=0.7\textwidth,angle=90,clip]{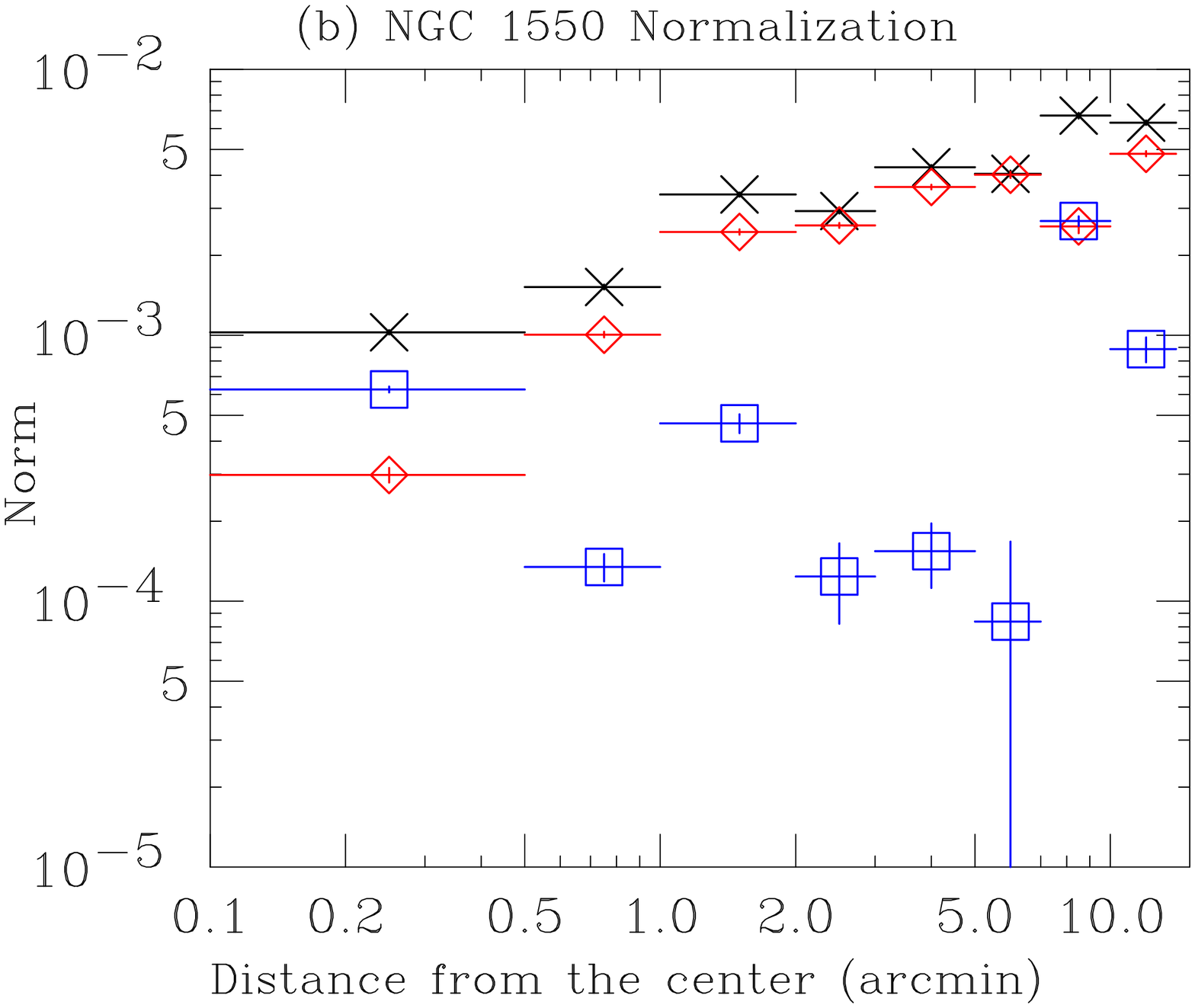}
  \end{center}
  \end{minipage}
  \begin{minipage}{0.5\hsize}
\vspace{0.5cm}
  \begin{center}
\includegraphics[width=0.7\textwidth,angle=90,clip]{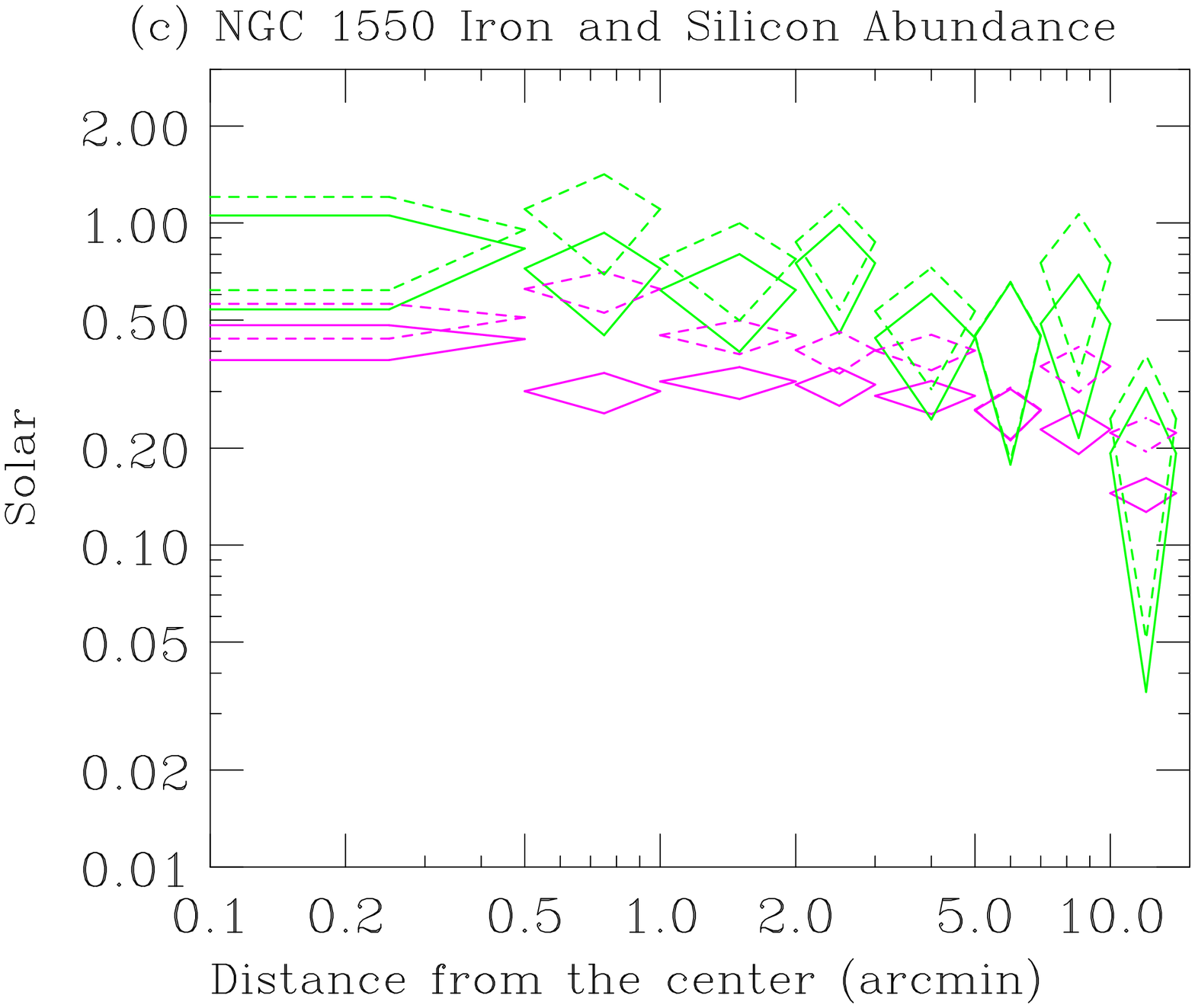}
  \end{center}
  \end{minipage}
  \begin{minipage}{0.5\hsize}
\vspace{0.5cm}
  \begin{center}
\includegraphics[width=0.7\textwidth,angle=90,clip]{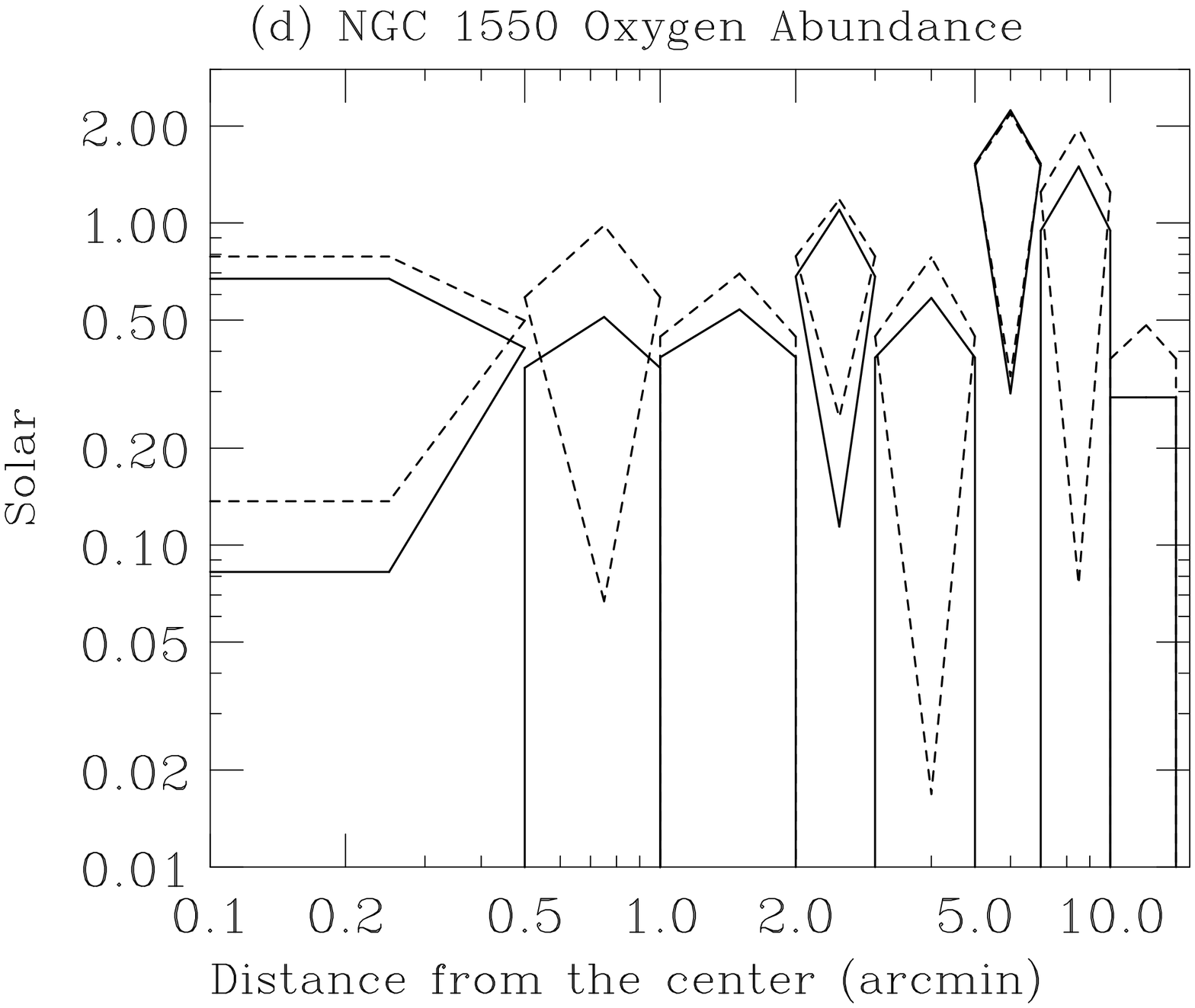}
  \end{center}
  \end{minipage}
  \begin{minipage}{0.5\hsize}
\vspace{0.5cm}
  \begin{center}
\includegraphics[width=0.7\textwidth,angle=90,clip]{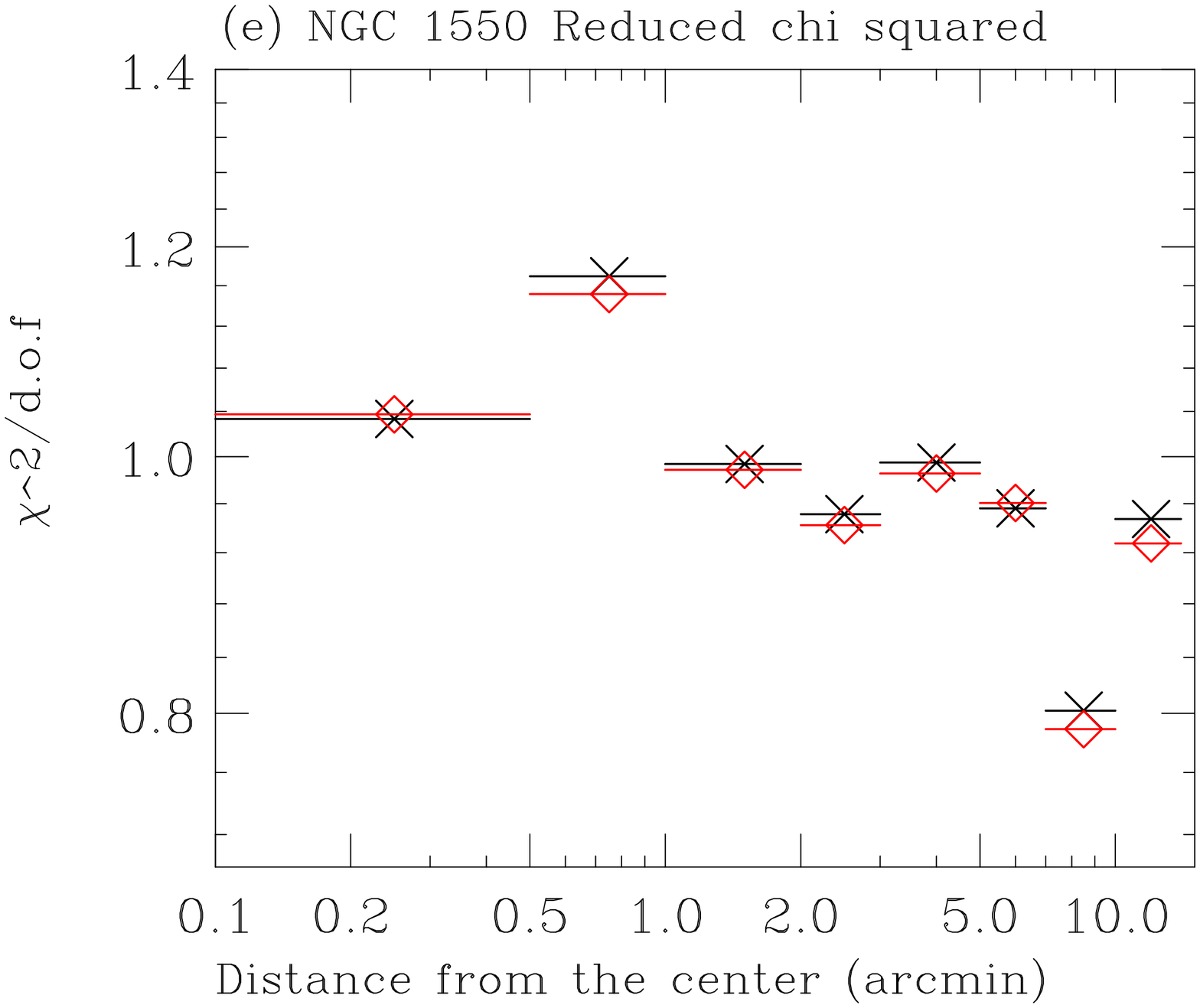}
  \end{center}
  \end{minipage}

  \caption{The same as Figure~\ref{fig:annular-prof}, but for the deprojected spectra.}
\label{fig:shell-prof}
\end{figure}

\begin{figure}
  \begin{center}
    %%% \FigureFile(width,height){filename}
%\includegraphics[width=0.4\textwidth,angle=0,clip]{./fig/2mass_prof/n1550_congra_img.eps}
\includegraphics[width=0.35\textwidth,angle=90,clip]{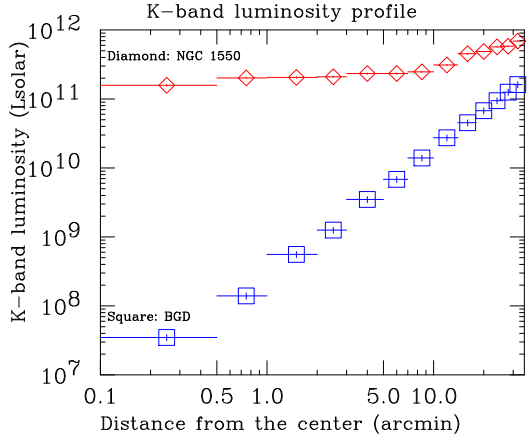}
  \end{center}
  \caption{
%(left) Positions of the 33 galaxies in the 2MASS catalog, found within $34'$
%of NGC~1550, plotted on J2000 ($\alpha$, $\delta$) coordinates. 
%Their serial numbers are also given. Colors specify their $K$-band 
%magnitudes as $8<K<9$ in green, $9<K<10$ in blue, $10<K<11$ in yellow, 
%$11<K<12$ in magenta, $12<K<13$ in cyan, and $13<K$ in black. 
%The contours show 0.5--10.0 keV MOS1 image, presented in the
%same manner as Figure~\ref{fig:n1550_image}.
A two dimensionally integrated $K$-band luminosity profile 
around NGC~1550 (red diamonds), compared with that of the estimated 
field galaxies (blue squares). The latter is calculated assuming 
that all the galaxies are at the same distance as NGC~1550. 
}
\label{fig:k-lumi_prof}
\end{figure}

\begin{figure}
  \begin{center}
\includegraphics[width=0.35\textwidth,angle=90,clip]{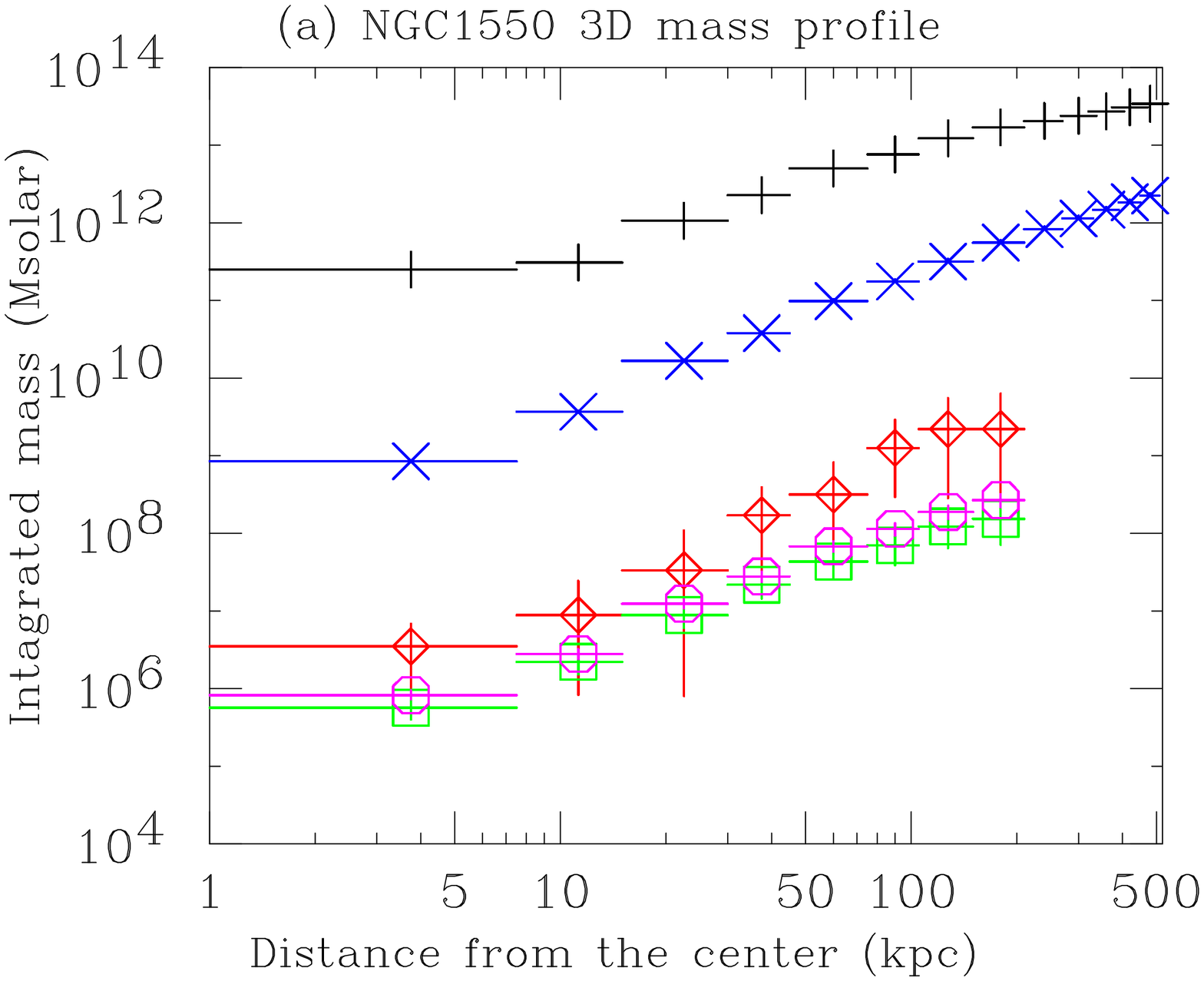}
\includegraphics[width=0.35\textwidth,angle=90,clip]{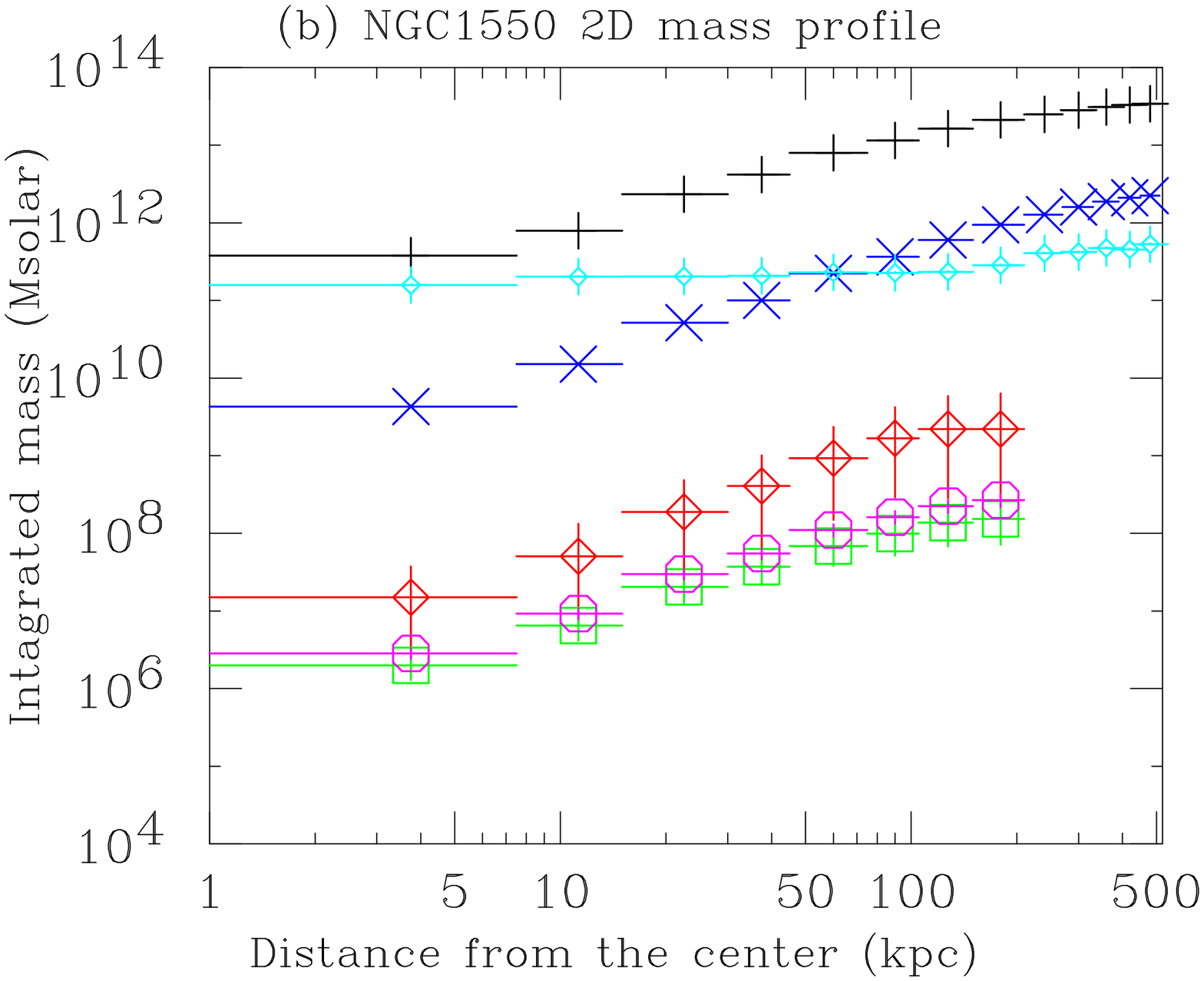}
  \end{center}
  \caption{
Integrated radial profiles of various mass components in the
NGC~1550 system, shown in 3-dimensional forms (panel a) and projected forms
(panel b). Black crosses indicate the total mass calculated by 
equation~(\ref{eqn:tot-mass}), while blue crosses the total gas mass obtained 
by equation~(\ref{eqn:gas-mass}).
Small cyan diamonds are the galaxy mass derived in \S 4, shown only in the
projected form. Metals in the gas are specified by colors;
O (red diamonds), Si (green squares), and Fe (magenta circles).
}
\label{fig:mass-prof}
\end{figure}

Using the {\it XMM-Newton} EPIC and 2MASS $K$-band data, we derived 
the radial profiles of components constituting the NGC~1550 system; 
the gas mass, total mass,  metal mass, and the galaxy luminosity.
As seen in Figure~\ref{fig:shell-prof}c and Figure~\ref{fig:mass-prof}, 
the metals (oxygen, silicon, and iron) are extended to $\sim 200$ kpc
from the center, wherein $\sim$ 70\% of 
the $K$-band luminosity is carried by NGC~1550 itself.
Most importantly, we have found that iron (as well as silicon)
in the ICM traces very well the total gravitating mass 
(eq.~[\ref{eqn:tot-mass}]), whereas the stellar component
is significantly more concentrated to $r < 15$ kpc.
This is a non-trivial result, since we would naturally expect
the metals to trace the stellar mass unless at least either
of them evolved significantly over the Hubble time in the
radial distribution. 

The above finding allows three alternative scenarios; 
one is that the iron within $\sim 100$ kpc has been transported 
outwards; another is that iron in the central region is contained in 
the central galaxy and hidden from our X-ray view. The other is that 
the galaxies have gradually concentrated toward the center 
(particularly onto NGC~1550) while supplying metals to the 
surrounding hot gas. 

\begin{figure}

\begin{minipage}{0.5\hsize}
%\vspace{0.5cm}
  \begin{center}
\includegraphics[width=0.7\textwidth,angle=90,clip]{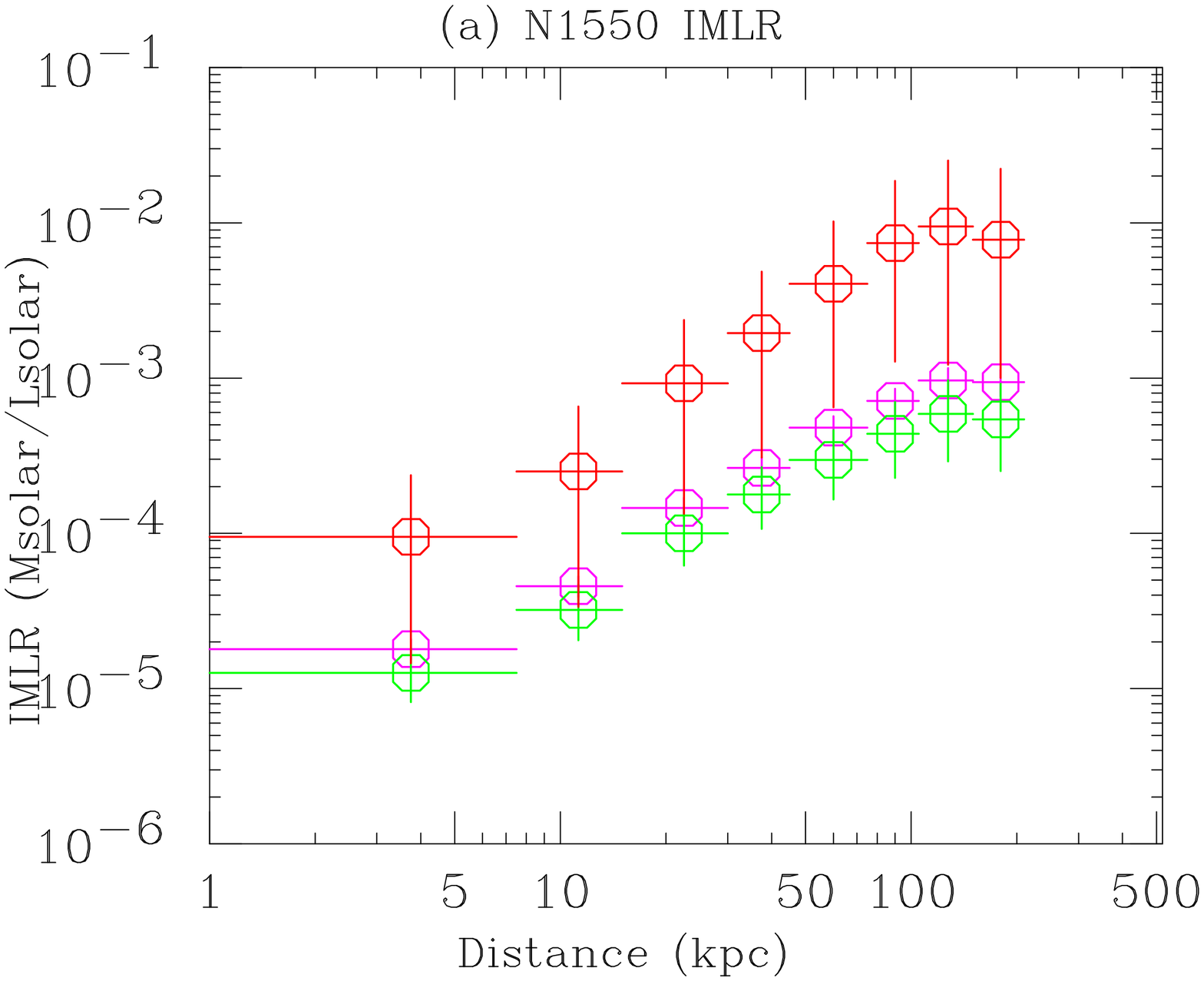}
 \end{center}
  \end{minipage}
\begin{minipage}{0.5\hsize}
%\vspace{0.5cm}
  \begin{center}
\includegraphics[width=0.75\textwidth,angle=0,clip]{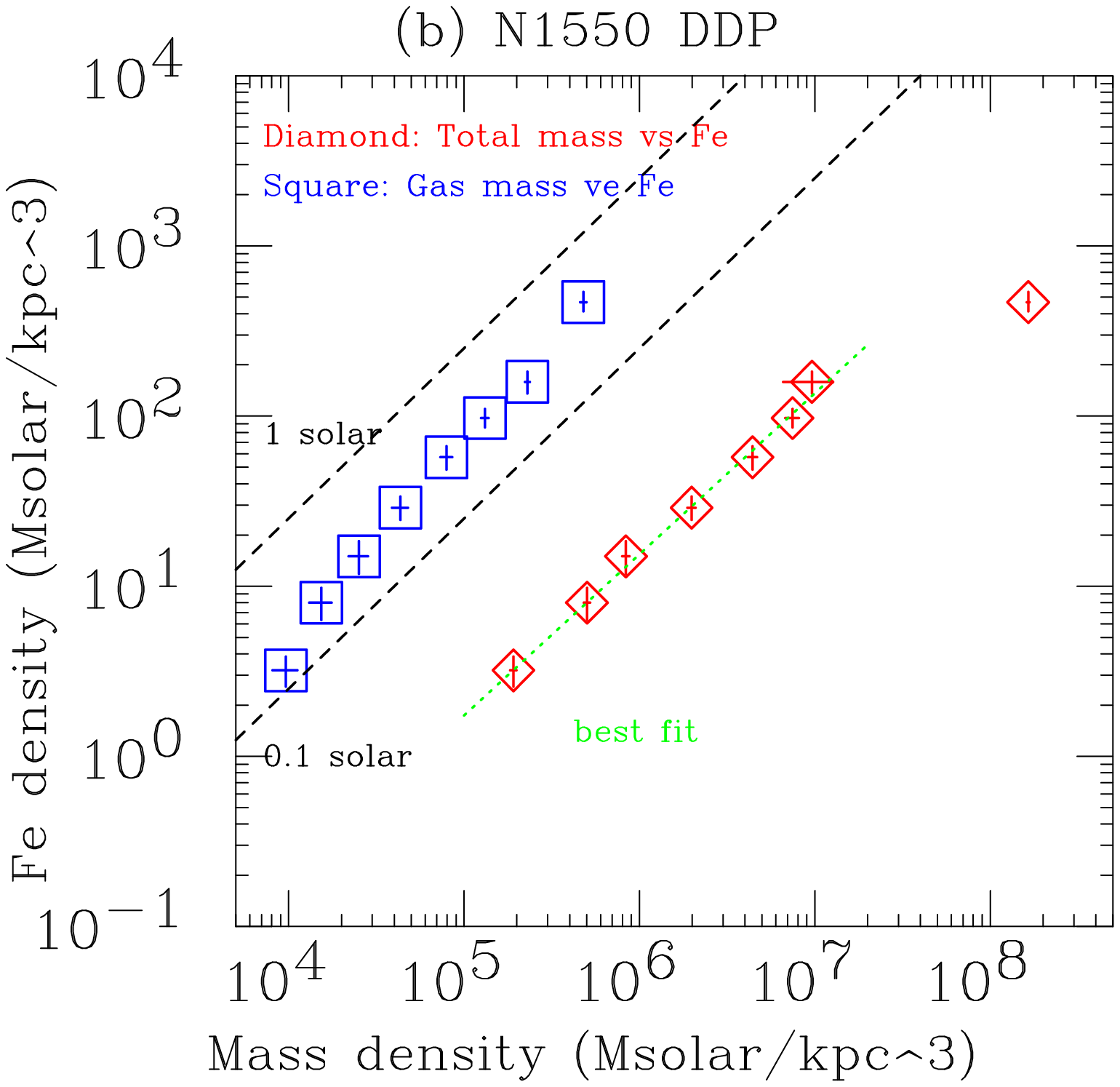}
  \end{center}
  \end{minipage}
\begin{minipage}{0.5\hsize}
\vspace{0.5cm}
  \begin{center}
\includegraphics[width=0.75\textwidth,angle=0,clip]{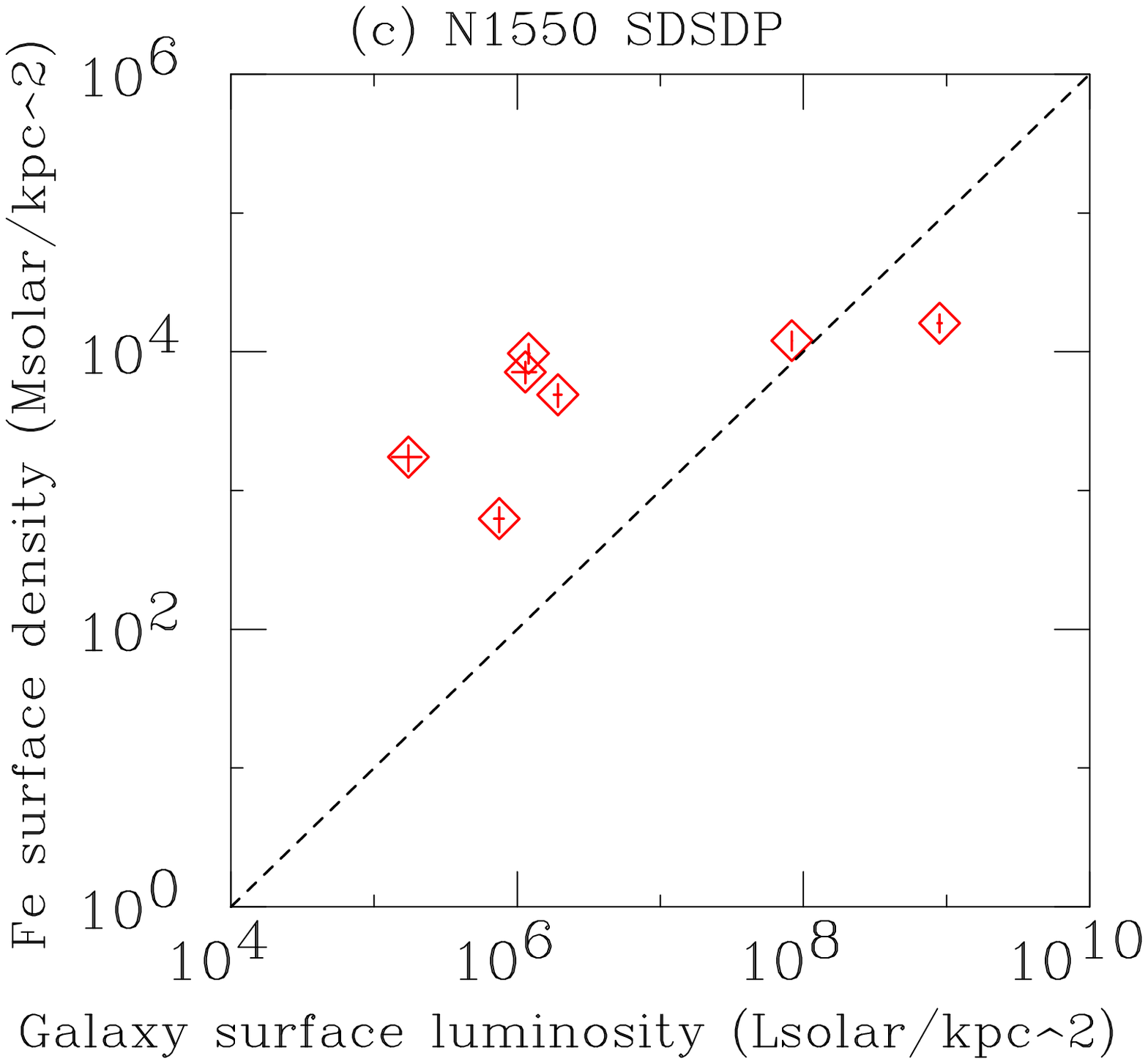}
 \end{center}
  \end{minipage}

%  \end{center}
  \caption{
(a) Radially-integrated
$K$-band iron-mass-to-light (magenta), silicon-mass-to-light (green), and
oxygen-mass-to-light (red) ratio profiles of NGC~1550. 
For convenience, 
the angular separation is converted to physical distance, assuming
that all the galaxies are at the same distance as NGC~1550.
(b) Density-density plots between the total mass and iron mass (red diamonds),
and between the gas mass and iron mass (blue squares).
Two dashed lines indicate 0.1 and 1.0 solar abundances of the iron mass.
Green dotted line shows the best fit power-law relation between 
the iron density and the total mass density, $\rho_{\rm Fe} = 3.191 \times 10^{-5}\; \rho_{\rm tot}^{0.947}$, excluding the innermost data point. 
(c) A surface density-surface density plot between the galaxy $K$-band 
luminosity and iron mass (red diamonds). The dashed line represents 
$\Sigma_{\rm Fe} \propto \Sigma_{\rm gal}$ for reference.
}
\label{fig:imlr-ddp}
\end{figure}

In the first scenario, metals in the central region must have been 
transported outward, together with the ICM (mostly hydrogen and helium),   
through diffusion or bulk outflow. As discussed in 
Ezawa et al. (1997; 2001), the diffusion constant $D$ of iron 
in the hot gas is given as
\begin{equation}
D = 2 \times 10^{26} \left( \frac{n_{\rm e}}{10^{-4}\; {\rm cm}^{-3}}  \right)^{-1} \left( \frac{kT}{3\; {\rm keV}}  \right)^{5/2} {\rm cm}^2 \; {\rm s}^{-1}. 
\end{equation}
Therefore, the iron atoms in the ICM can diffuse 
at most $\sim 10$ kpc over the Hubble time; this is by far insufficient
to explain the present results. 
In the case of bulk outflow, a large amount of hot gas within 100 kpc, 
comparable in mass to what presently exists there, must have been 
evacuated to realize the IMLR of Figure~\ref{fig:imlr-ddp}a
and SDSDS of Figure~\ref{fig:imlr-ddp}c.
Since the mass of the hot gas within $100$ kpc is 
$6 \times 10^{11} M_{\odot}$ (Figure~\ref{fig:mass-prof}b),  
the outflow rate becomes $\sim 50\; M_{\odot} {\rm yr}^{-1}$ if we assume 
a constant rate over the Hubble time.
This is just opposite to the case of cooling flows. In other words, 
the energy needed to drive out the central ICM must have been 
large enough to reverse cooling flows. However, even among 
medium-richness clusters, it is generally considered non-trivial to
find an energy source which is strong enough to suppress cooling flows
of this order (e.g. Makishima et al. 2001). 
Since the depth of gravitational potential is comparable
to the velocity dispersion of galaxies $\sigma_{\rm gal}$, 
the energy required to evacuate the central ICM, $E_{\rm eva}$, 
becomes enormous, as
\begin{eqnarray}
E_{\rm eva} \sim M_{\rm eva}  \sigma_{\rm gal}^2 = 3 \times 10^{60} \left(\frac
{\sigma_{\rm gal}}{400\; {\rm km}\; {\rm s}^{-1}}\right)^2 \; {\rm erg}. 
\label{eqn:evacuation}
\end{eqnarray}
where $M_{\rm eva} \sim 1 \times\; 10^{12} M_{\odot}$ is a typical ICM mass 
to be evacuated.

The most popular candidate for driving the outflow is
activity of the central galaxy nucleus.
Actually, NGC~1550 hosts a weak radio source of which the flux density is
$16.6 \pm 1.6$ mJy at 1.4 GHz (Condon et al. 1998).
However, for the estimated kinetic energy (power times age) 
provided by an AGN to become comparable to equation~(\ref{eqn:evacuation}),
the object needs to be as strong a radio source as Cygnus A 
(e.g. Kino et al. 2005); the present radio luminosity of the NGC~1550
nucleus falls 6 orders of magnitude in short. Of course, we cannot
exclude an extreme possibility that the nucleus remained highly active
(either continuously or intermittently) till a near past, and then
declined at present. In such a case, we would observe large 
($\sim 100$ kpc) cavities in the ICM, but there is no such evidence. 
Alternatively, the radiation pressure from the AGN may have
caused the outflow. If we assume the ICM within 100 kpc
from the center has been pushed away, the luminosity of the AGN has to be 
\begin{eqnarray}
L_{\rm AGN} = \left(\frac{\dot{E}_{\rm eva}}{\tau_{\rm T}}\right) = 4 \times 10^{46} \; {\rm erg}\; {\rm s}^{-1}, 
\end{eqnarray}
where $\dot{E}_{\rm eva}$ is the evacuation power averaged over the Hubble time,
and  $\tau_{\rm T} = 2 \times 10^{-4}$ is the cross section 
for Thomson scattering.
It is unlikely that the NGC~1550 nucleus has been so luminous for 
the Hubble time.
Yet another weak point of the scenario is that it needs some 
fine tuning mechanisms to make $\rho_{\rm Fe}$ proportional to 
$\rho_{\rm tot}$ as indicated by equation~(\ref{eqn:tot-mass}) and 
the DDP of Figure~\ref{fig:imlr-ddp}b. Thus, the bulk outflow
scenario is also unlikely.

The second scenario assumes that the IMLR decreases toward the
center because a significant fraction of iron in the central
region has been taken inside the stars in NGC~1550, thus
increasing the stellar metallicity of NGC~1550 compared to
those of the other peripheral galaxies.
This ``hidden''amount of iron becomes comparable to that in the hot gas
within 100 kpc, that is, $2 \times 10^{8} M_{\odot}$ 
(Figure~\ref{fig:mass-prof}b). For comparison, the stellar
mass in NGC~1550 is estimated to be $2 \times 10^{11} M_{\odot}$, 
assuming again the $K$-band stellar mass-to-light ratio to be 1. 
Therefore, the hidden iron should increase the mean stellar metal abundance
by $\sim 0.5$ in solar units. In this estimate, we can ignore
the amount of iron confined in interstellar matter (ISM) in these
galaxies, because the mass of ISM is at most 10\% of the stellar mass 
(Matsushita 2000). From a color-metallicity relation studied by
Barmby et al. (2000), the stellar metallicity of NGC~1550 is estimated to
be $0.5-0.7$ solar abundance, using $U$,$B$,$V$,$J$, and $K$ magnitude. 
On the other hand, Jord\'an et al. (2004) showed that the metallicity 
of early-type galaxies correlates well with the absolute magnitude
$M_V$. From this correlation, the metallicity of galaxies having the same
$M_V$ as NGC~1550 is estimated as $\sim 0.8$ solar. 
Since this is not much different from the value estimated from
the color of NGC~1550, there is no reason to presume that NGC~1550 has 
locked a much larger fraction of its metal products into its 
stars than other elliptical galaxies with simillar 
$V$-band luminosities. 

The remaining scenario assumes gradual concentration, or infall, of 
the member galaxies toward the system center. This view is suggested by 
the present and other (e.g., Makishima et al. 2001) results, that the 
stellar mass in groups and clusters are generally much more concentrated,
not only than the metals in the ICM, but also than the ICM itself
and the total mass (Figure~\ref{fig:mass-prof}b). The former two scenarios 
would not easily explain these differences among the metals, stars,
and the total mass (dominated by the dark matter). To be more quantitative,
let us assume, as a simplest case, that the system initially consisted 
mainly of $\sim 10$ galaxies each with $K \sim 11$, which is comparable 
to IC~0366 but an order of magnitude less luminous than the present-day 
NGC~1550. If only one of them was initially at $ r \lesssim 50$ kpc 
while the others were outside, the IMLR at this radius was initially 
an order of magnitude higher than it is at present, and hence the 
IMLR profile at that time was approximately flat as would be naturally 
expected. If these galaxies have all merged into the central single galaxy 
(i.e., the present-day NGC~1550) after their metal supply was mostly over, 
we can explain both its present optical luminosity, and the central 
decrease in the IMLR profile. 
In other words, NGC~1550 is indeed suggested to be a remnant of such 
a merger process, as implied by the nomenclature of ``fossil group" 
(Jones et al. 2003).

As to the mechanism which drives the suggested galaxy infall,
a promising candidate is dynamical pressure  exerted onto galaxies 
when they swim through the ICM. According to Sarazin (1988),
this effect will cause the kinetic energy $K$ of a galaxy,
moving  with a velocity $v$, to decrease at a rate of
\begin{equation}
dK/dt = -\pi \rho_{\rm gas} v^3 R_{\rm int}^2~~,
\label{eq:interaction}
\end{equation}
where $R_{\rm int}$ is ``interaction radius''
with which the galaxy interacts with the ICM and displaces it.
Then, the galaxy will lose
its kinetic energy on a time scale of
\begin{equation}
\tau \equiv  \frac{K}{ |dK/dt|} = \frac{m_{\rm gal}}{2\pi \rho_{\rm gas} v R_{\rm int}^2} ~~ ,
\label{eq:timescale1}
\end{equation}
where $m_{\rm gal}$ is the galaxy mass.

As discussed by Makishima et al. (2001),
the value of $R_{\rm int}$ is estimated
to be comparable to the visual galaxy size,
because the in-flowing ICM will be intercepted
by magnetic fields anchored to the galaxy.
Assuming hence $R_{\rm int}=10$ kpc,
and employing plausible values for the other parameters,
equation (\ref{eq:timescale1}) yields
\begin{equation}
\tau  = 14 \left( \frac{m_{\rm gal}} {1\times 10^{11}M{_\odot}}   \right)
            \left( \frac{n_{\rm gas}} {10^{-3}\;{\rm cm^{-3}}  }   \right)^{-1}
            \left( \frac{v}{400\;{\rm km\;s^{-1}}      }   \right)^{-1}
            \left( \frac{R_{\rm int}}{10\;{\rm kpc}}      \right)^{-2}~{\rm Gyr}
\label{eq:timescale2}
\end{equation}
which is of the same order as the Hubble time.
Then, the galaxies moving in dense ICM will
fall toward the center, merge together,
and  change their morphology (Makishima et al. 2001),
on the Hubble time scale.
Importantly,  equation (\ref{eq:timescale2}) predicts
less massive  galaxies to fall more rapidly
(on condition that $R_{\rm int}$ depends only weakly on $m_{\rm gal}$),
in agreement with the idea
that small galaxies  which may have existed in the system
have mostly merged into NGC~1550.
This mass dependence is opposite to
that of the gravitational dynamical friction,
which works among galaxies moving in vacuum
and causes  more massive galaxies to sink to the potential center.

The scenario described above is not considered
specific to the NGC~1550 system,
since the dynamical pressure cannot be avoided
as long as galaxies are moving through the ICM.
Employing the empirical relation among the cluster mass $M$, the
X-ray luminosity $L_{\rm x}$, and the temperature $T$ as 
$L_{\rm x} \propto T^3$ and $M \propto T^{3/2}$ (e.g. Arnaud et al. 2005),
The ICM density $n_{\rm gas}$ in a cluster is found to scale 
as $\propto T^{1/2}$; here we also utilized the virial relation as
$M/R \propto T$, and the luminosity relation as 
$L_{\rm x} \propto n_{\rm gas}^2 R^3T^{1/2}$, where $R$ is the cluster radius.
Then,  with $v \propto T^{1/2}$,
equation (\ref{eq:timescale1}) predicts $\tau \propto T^{-1}$,
so that richer systems are expected to have
somewhat shorter merging time scales.
In fact, the central decrease in the IMLR profile has been observed
from richer systems as well (Finoguenov et al. 2000; Makishima et al. 2001; 
Kawaharada 2006), including the Centaurus cluster and Abell 1060.
Interestingly, in Figure~9 of Finoguenov et al. (2000), the IMLR profiles 
in HCG~51 and HCG~62 are flatter than those of the other groups and clusters.
This support the galaxy-infall scenario indirectly, because
a compact group is considered to be a galaxy group in which 
galaxy mergers are proceeding.
Furthermore, the ICM dynamical pressure  might also explain
some evolutionary effects observed from rich clusters,
such as the long-known Butcher-O\"emler effect (Butcher \& O\"emler 1978),
and the significant evolution of galaxy morphology in
dense cluster environments (e.g., van Dokkum et al. 2000; Goto 2005).

It is known that galaxy groups exhibit rather large scatter 
in their gas richness (Osmond \& Ponman 2004), with spiral-rich 
ones tending to be particularly gas poor (Ota et al 2004).
 From equation (\ref{eq:timescale2}), we may then speculate
that a galaxy group borne as a relatively gas-rich system
is subject to a fast merger process,
while it remains otherwise an ordinary group.
If this view is correct, an IXEGs could be a system
in which baryons remained for some reason mostly
in the hot gas phase,
with a relatively lower galaxy mass.
The NGC~1550 system agrees with this view, because it is as
gas-rich as ordinary galaxy groups in terms of the gas-to-total mass
ratio (Paper I), while it is optically comparable to a single galaxy.

\acknowledgments

We acknowledge support from the grant for young researchers 
from Japan Society for the Promotion of Science (JSPS), and the grant 
from Special Postdoctoral Researchers Program of RIKEN. 

{\it Facilities:} \facility{FLWO:2MASS}, \facility{XMM}.

\clearpage

%%######################### table 1 #######################################
\begin{deluxetable}{lcccccc}
\tabletypesize{\scriptsize}
%\rotate
\tablecaption{Results of the model fitting to the MOS and PN spectra within $14'$ of the NGC~1550 nucleus. \label{table:fit-result}}
\tablewidth{0pt}
\startdata
\hline %-----------------------------------------------------------------
\hline %-----------------------------------------------------------------
 & & & \multicolumn{3}{c}{abundances (solar)} & \\
\cline{4-6}
Model  &Temperature (keV)& Ratio\tablenotemark{a} & Fe & Si & O & $\chi^2/$dof   \\
\hline %-----------------------------------------------------------------
vMEKAL     &$1.32 \pm 0.01$ & -- & $0.29 \pm 0.01 $& $0.31 \pm 0.02 $ & $0.13 \pm 0.03$  & 2012/1423 = 1.41 \\
vAPEC     &$1.27 \pm 0.01$ & -- & $0.24 \pm 0.01$ & $0.32 \pm 0.02$ & $0.13 \pm 0.03 $ &1813/1425 = 1.27 \\
2vMEKAL     &$1.50 \pm 0.02$ / $0.87 \pm 0.01$ & 5.91 & $0.42 \pm 0.01$& $0.45 \pm 0.02$& $0.24 \pm 0.04 $& 1640/1421 = 1.15 \\
2vAPEC    & $1.48 \pm 0.02 $ / $0.95 \pm 0.01$ & 4.08 & $0.34 \pm 0.01$ &$0.40 \pm 0.02$ &$0.18 \pm 0.03$& 1666/1423 = 1.17 \\
%\hline %-----------------------------------------------------------------
\enddata
%% Text for table notes should follow after the \enddata but before
%% the \end{deluxetable}. Make sure there is at least one \tablenotemark
%% in the table for each \tablenotetext.
%\tablecomments{Using vAPEC model within $14'$ of the NGC~1550 nucleus}
\tablenotetext{a}{Ratio of the hot-component normalization to that of the cool component}
\end{deluxetable}
%%########################################################################

%%######################### table 2 #######################################
\begin{table}
  \caption{Major candidate of member galaxies in the NGC~1550 system.}
\label{table:member}
  \begin{center}
    \begin{tabular}{lcccccc}
\hline %-----------------------------------------------------------------
\hline %-----------------------------------------------------------------
Name & $\Delta \theta$ (arcmin) & $K$ (mag)& $v_r$ (km s$^{-1}$)& \\
\hline %-----------------------------------------------------------------
NGC~1550 & 0 & 8.774 & 3714   \\
IC~0366 & 3.1 & 11.104 & 3736 \\
UGC~03011 & 12.0 & 10.891 & 4207 \\
MCG+00-11-050 & 12.7 & 10.840 & -- \\
UGC~03008 & 17.0 & 9.687 & 3251 \\
CGCG~393-007  & 17.1 & 10.308 & 3912 \\
CGCG~393-008 & 23.6 & 9.955 & 3973 \\
UGC~03006 & 33.2 & 9.581 & 3664 \\
\hline %-----------------------------------------------------------------
    \end{tabular}
  \end{center}
\end{table}
%%########################################################################

\clearpage

%%#################### figure 1 #########################################
%%%#######################################################################

%%#################### figure 2 #########################################
%%%#######################################################################

%%#################### figure 3 #########################################
%%#######################################################################

%%#################### figure 4 #########################################
%%#######################################################################

%%#################### figure 5 #########################################
%%%#######################################################################

%%%#################### figure 6 #########################################
%\begin{figure}
%  \begin{center}
%%     \FigureFile(80mm,80mm){./fig/0-14_fit/n1550_0-14_sys.ps}
%    \includegraphics[width=0.38\textwidth,angle=-90,clip]{./fig/epic_sys_kt.eps}  
 %   \includegraphics[width=0.38\textwidth,angle=-90,clip]{./fig/epic_sys_metal.eps}  
 % \end{center}
 % \caption{(left) Temperatures determined with $0.5-5.0$ keV PN, 
%MOS1, and MOS2 spectra extracted from the $0'-14'$ region, fitted
%separately using the two-temperature vAPEC model. The metal abundances 
%are tied between the two components, while the temperatures and 
%normalizations are left free.
%(right) the same as left panel, but for the iron and oxygen abundances.
%}
%\label{fig:spec-sys}
%\end{figure}%
%%%#######################################################################

%%#################### figure 6 #########################################
%%#######################################################################
%%#################### figure 7 #########################################
%%#######################################################################

%%#################### figure 8 #########################################
%%#######################################################################

%%#################### figure 9 #########################################
%%#######################################################################

%%%#################### figure 10 #########################################
%%%#######################################################################

\end{document}